\documentclass[review]{elsarticle}
\usepackage{amsmath}
\usepackage{amsfonts}
\usepackage{amssymb}
\usepackage{graphicx}

\usepackage{indentfirst}

\begin{document}

\begin{frontmatter}

\title{Strong field double ionization of ${\rm H}_2$ : Insights from nonlinear dynamics}

\author[EC,CPT]{F. Mauger}
\author[CPT]{C. Chandre\corref{cor1}}
\ead{chandre@cpt.univ-mrs.fr}
\author[GA]{T. Uzer}
\cortext[cor1]{Corresponding author}
\address[EC]{Ecole Centrale de Marseille, Technop\^ole de Ch\^ateau-Gombert, 38 rue Fr\'ed\'eric Joliot Curie
F-13451 Marseille Cedex 20, France}
\address[CPT]{Centre de Physique Th\'eorique, CNRS -- Aix-Marseille Universit\'es, Campus de Luminy, case 907, F-13288 Marseille cedex 09, France}
\address[GA]{School of Physics, Georgia Institute of Technology, Atlanta, GA 30332-0430, USA}
\date{\today}

\begin{abstract}
The uncorrelated (``sequential'') and correlated (``nonsequential'') double ionization of the ${\rm H}_2$ molecule in strong laser pulses is investigated using the tools of nonlinear dynamics. We focus on the phase-space dynamics of this system, specifically by finding the dynamical structures that regulate these ionization processes. The emerging picture complements the recollision scenario by clarifying the distinct roles played by the recolliding and core electrons. Our analysis leads to verifiable predictions of the intensities where qualitative changes in ionization occur. We also show how these findings depend on the internuclear distance.  
\end{abstract}


\end{frontmatter}

\section{Introduction}

The interaction of strong fields with molecules and their molecular ion and isotopic species have been studied extensively (see Refs.~\cite{post04,band06u,klin08} for reviews). Among molecular species, the ${\rm H}_2$-laser interaction remains a subject of particular interest because of the attosecond and femtosecond nuclear time scales and the relatively simple decay channels, showing the way to controlling the motion of electrons in molecules~\cite{band04,chel06,klin06}. Nonsequential double ionization of ${\rm H}_2$ has been studied in detail with an emphasis on the rescattering dynamics (e.g.~\cite{tong03}). 

\begin{figure}
	\centering
		\includegraphics[width = \linewidth]{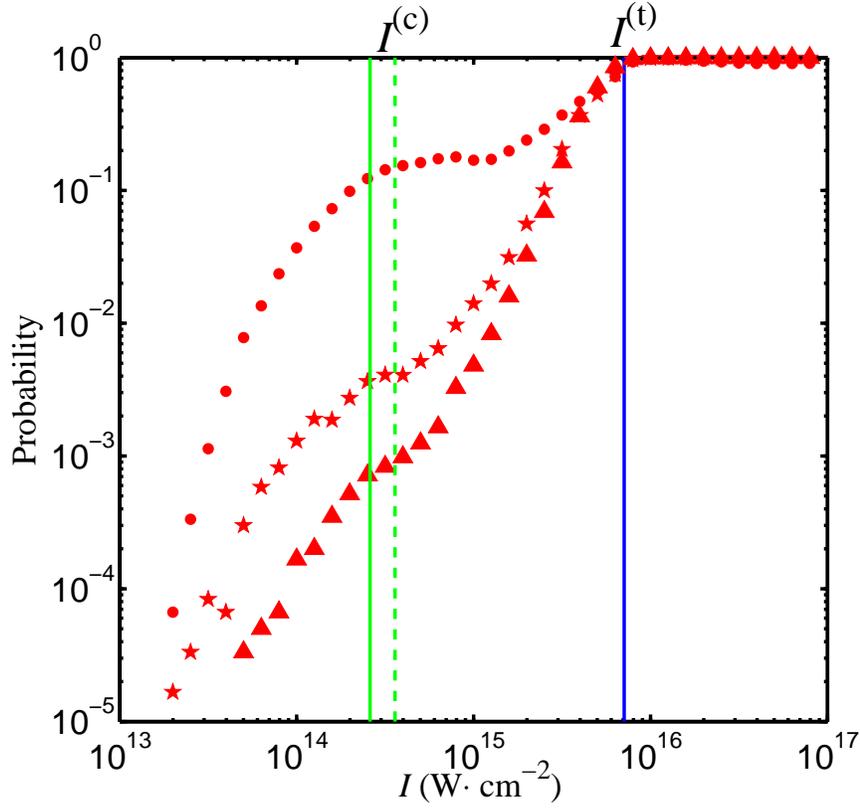}
	\caption{\label{fig:Fig1}
	Double ionization probability for Hamiltonian~(\ref{Eq:Hamiltonian}) for the~$d=1$ (dots)~$d=2$ (stars) and $d=3$ (triangles) models as a function of the intensity of the field~$I$ for~$\omega = 0.0584$.  The vertical lines indicate the laser intensities (continuous green line) $I^{(c)}\approx 3.60 \times 10^{14}\ \mbox{W} \cdot \mbox{cm}^{-2}$  and (dashed green line) $I^{(c)}\approx 3.60 \times 10^{14}\ \mbox{W} \cdot \mbox{cm}^{-2}$ where our dynamical analysis predicts the maximum of nonsequential double ionization (related to two different approximations), and (in blue) the intensity $I^{(t)}\approx 7.10 \times 10^{15}\ \mbox{W} \cdot \mbox{cm}^{-2}$ where the double ionization is expected to be complete.}
\end{figure}

The strong enhancement of nonsequential (correlated) double ionization over its sequential (uncorrelated) counterpart is one of the most striking surprises in laser-matter interactions. Indeed, at some intensities, double ionization rates can be 
several orders of magnitude higher than the uncorrelated sequential mechanism leads one to believe. 
The hallmark of this strong correlation is the so-called knee of the double ionization probability versus the laser intensity. An example of a knee is provided in Fig.~\ref{fig:Fig1} for a model of the ${\rm H}_2$ molecule. Similar knees have been observed in 
experimental data~\cite{fitti92,kodo93,walk94,laro98,webe00_2,corn00,guo01,dewi01,ruda04} and successfully 
reproduced by quantal computations on atoms and molecules~\cite{beck96,wats97,lapp98,panf03,Baie06}. Different scenarios 
have been proposed to explain the mechanism behind this enhancement~\cite{fitti92,Cork93,scha93,walk94,beck96,kopo00,lein00,sach01,fu01,panf01,barn03,colg04,ho05_1,ho05_2,ruiz05,horn07,prau07,feis08}. When confronted with experiments~\cite{brya06,webe00_2}, 
the recollision scenario~\cite{Cork93,scha93}, in which an ionized electron is hurled back at the core and ionizes the second electron, seems in best accord with observations. It is generally agreed that the sequential process stems from the action of the laser pulse, whereas the electron-electron correlation is at the essence of the nonsequential process. 

While classical mechanics would, at first sight, seem an inappropriate tool for studying this strong-field process, recent classical models~\cite{fu01,sach01,panf02,panf03,ho05_1,ho05_2,liu07,maug09} have been surprisingly successful in capturing key features of the process because of the dominant role of correlation~\cite{ho05_1}. Indeed, entirely classical interactions turn out to be adequate to generate the strong two-electron correlation needed for double ionization. These ideas have been put to the test on the helium atom~\cite{fu01,sach01,panf02,panf03,ho05_1,ho05_2,maug09}. A purely classical scenario for the double ionization of the Helium atom has been proposed in Ref.~\cite{maug09} based on the identification of the organizing centers of the classical dynamics of an effective Hamiltonian model. In the current manuscript, we explore the generality of this scenario and in particular how these ideas apply to the double ionization of the ${\rm H}_2$ molecule driven by a linearly polarized strong laser pulse. We shall show that the ideas developed fore helium are also applicable here, with some natural variations for the presence of the separated charge centers. In summary, the emerging classical scenario for nonsequential double ionization is:
 
\begin{itemize}
\item Without the field, the dynamics of ${\rm H}_2$ is very chaotic and mainly driven by four weakly hyperbolic periodic orbits. These periodic orbits define an ``inner'' and an ``outer'' electron.
\item At time $t=0$, the laser pulse is switched on and focuses its action on the outer electron by pulling it out of the core region. 
\item After half a period of the field, the outer electron is hurled back to the core region and interacts with the inner electron for a very short time. The interaction, defined as a recollision, transfers energy from the outer to the inner electron. Several outcomes are possible and among them are~: A double ionization where both electrons leave the inner region, a single ionization where one of the electron leaves the nucleus and the other one stays in its neighborhood, an exchange of roles between the inner and the outer electron, or the outer electron proceeding for another recollision.
\item Between two recollisions, the inner and outer electrons have their own respective dynamics. The inner region is mainly influenced by a periodic orbit which has the same period as the laser field. As long as this periodic orbit exists, it organizes an inner region where all the trajectories of the inner electron remain bounded. 
\end{itemize}

This scenario complements the recollision scenario by adding the phase space picture of the inner electron to the one of the outer electron. The mechanism in phase space allows us to formulate some verifiable predictions of two characteristic intensities of the knee~: The intensity $I^{(c)}$ where the nonsequential double ionization is expected to be maximum and the intensity $I^{(t)}$ where the double ionization is complete (see Fig.~\ref{fig:Fig1}). In comparison with the helium case, we highlight the common features but also the differences generated by the two nuclei, and in particular how the predictions depend on the distance between the two nuclei.
 
\section{Hamiltonian models for ${\rm H}_2$}

\begin{figure}
	\centering
	\includegraphics[width = \linewidth]{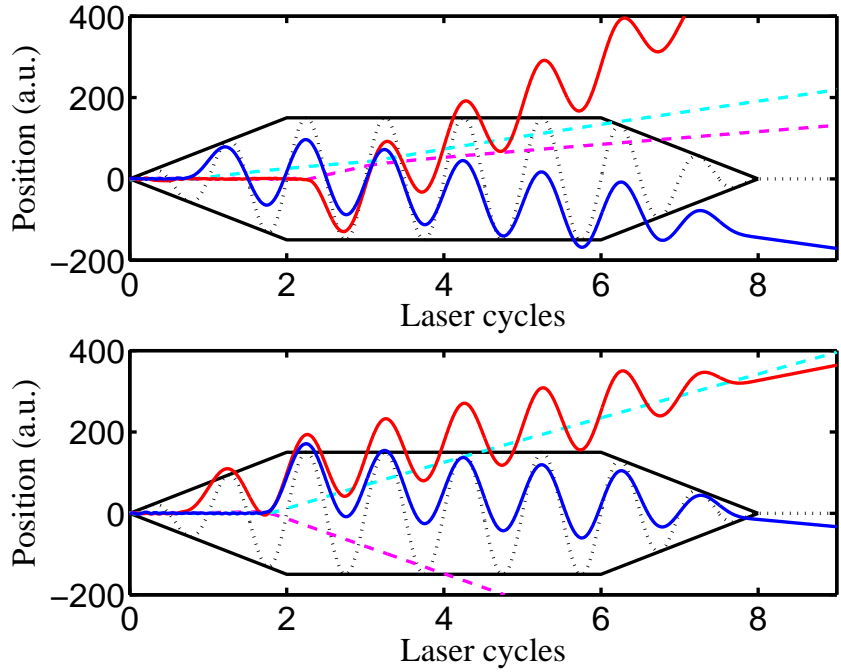}
	\caption{\label{fig:Fig2}
	Example of sequential (upper panel) and nonsequential (lower panel) double ionization trajectories of Hamiltonian~(\ref{Eq:Hamiltonian})~(for the $d=2$ model) for~$\omega = 0.0584$ and~$I = 3 \times 10^{15}\ \mbox{W} \cdot \mbox{cm}^{-2}$. The positions along the polarization axis $x_1$ (continuous blue curve) and $y_1$ (continuous red curve) of each pair of electrons are plotted versus time, as well as the positions in the transverse direction $x_2$ (dashed cyan curve) and $y_2$ (dashed magenta curve). The pulse shape function~$f(t)$ (black curve) and the laser excitation~$E(t)$~(dotted black curve) are also represented. The amplitude of the shape function and of the laser excitation are not representative of the actual conditions.}
\end{figure}

We consider a model for ${\rm H}_2$ where the two nuclei are fixed~\cite{Prau05,Baie06}, motivated by the fact that the time scale of electron motion during the laser pulse is short compared to the one characterizing the motion of the heavy nuclei. More specifically, we consider the following Hamiltonian model of the~$H_{2}$ molecule (in atomic units) with soft Coulomb potentials~\cite{Prau05,Baie06, Lein02, Saug08}~:
\begin{eqnarray} \label{Eq:Hamiltonian}
   \mathcal{H} \left( \mathbf{x}, \mathbf{y}, \mathbf{p}_{x}, \mathbf{p}_{y}, t \right) =
      \frac{\left\| \mathbf{p}_{x} \right\|^{2}}{2} 
      + \frac{\left\| \mathbf{p}_{y} \right\|^{2}}{2} 
      + \frac{1}{\left\| \mathbf{R} \right\|} \nonumber \\ 
      + \frac{1}{\sqrt{\left\| \mathbf{x} - \mathbf{y} \right\|^{2} + 1}} 
      + \left( \mathbf{x} + \mathbf{y} \right) \cdot \mathbf{E}\left( t \right) \nonumber \\
      -\frac{1}{\sqrt{\left\|{\bf R}/2 - \mathbf{x} \right\|^{2} + 1}}
      -\frac{1}{\sqrt{\left\| {\bf R}/2 + \mathbf{x} \right\|^{2} + 1}} \nonumber \\
      -\frac{1}{\sqrt{\left\| {\bf R}/2 - \mathbf{y} \right\|^{2} + 1}}
      -\frac{1}{\sqrt{\left\| {\bf R}/2 + \mathbf{y} \right\|^{2} + 1}},
\end{eqnarray}
where~$\mathbf{x}$ and $\mathbf{y}$ are the positions of the two electrons in a $d$-dimensional space,  and~$\mathbf{p}_{x}$,~$\mathbf{p}_{y}$ are their canonically conjugate momenta, and $\cdot$ denotes the Euclidean scalar product in $\mathbb{R}^d$ and $\left\| \cdot \right\|$ its corresponding norm. Here we consider $d=1$, 2 and 3 for the computations, but we mainly consider $d=1$ for the phase space analysis since we will see below that it reproduces the same characteristic features as the higher dimensional cases. The energy is initially fixed at the ground state~$\mathcal{E}_{g} = -1.16 \mbox{ a.u.}$~\cite{Lein02, Saug08}. The linearly polarized laser field is a sinusoidal pulse modulated by an envelope,~i.e.~$\mathbf{E}(t) = \mathbf{e}_1 E_{0} f(t) \sin \omega t$ where $\bf{e}_1$ is a unit vector characterizing the direction of the polarization, $E_{0}$ is the maximum amplitude and~$\omega$ the laser frequency chosen at~$\omega = 0.0584 \mbox{ a.u.}$ which corresponds to a wavelength of $780$~nm. The pulse envelope~$f(t)$ is chosen as a trapezoidal function with 2-4-2~laser pulse shape (the ramp-up lasts two cycles, the plateau four and the ramp-down two). We assume the two nuclei are aligned with the laser field and the distance, denoted~$R$, between them is fixed during the whole range of time we will consider~\cite{Baie06}, so that~$\mathbf{R} = R \mathbf{e}_1$ where $R=1.4 \mbox{ a.u.}$~\cite{Saug08}. \\
Typical ionizing trajectories of Hamiltonian~(\ref{Eq:Hamiltonian}) show mainly two qualitatively different routes to double ionization~(see Fig.~\ref{fig:Fig2})~: nonsequential double ionization~(NSDI), where the two electrons leave the core~(inner) region at about the same time, and sequential double ionization~(SDI), where one electron leaves the inner region long after the other one has ionized.

Using a large assembly of initial conditions, we compute the double ionization probability as a function of the laser intensity $I$ (related to $E_0$ by $I(\mbox{W} \cdot \mbox{cm}^{-2})=3.521 \times 10^{16} E_0({\rm a.u.})^2$) for $d=1$, 2 and 3 (see Fig.~\ref{fig:Fig1}). This probability curve has the shape of a knee, a hallmark of double ionization. 
Even if the probability of double ionization is much lower for the three or the two dimensional model, one can see that characteristic features (the localization of the knee and the intensity where double ionization is complete) happen at about the same intensities. We restrict ourselves to the one-dimensional case in what follows in order to obtain some information on these intensities and the mechanisms behind NSDI and SDI.

\begin{figure}
	\centering
		\includegraphics[width = \linewidth]{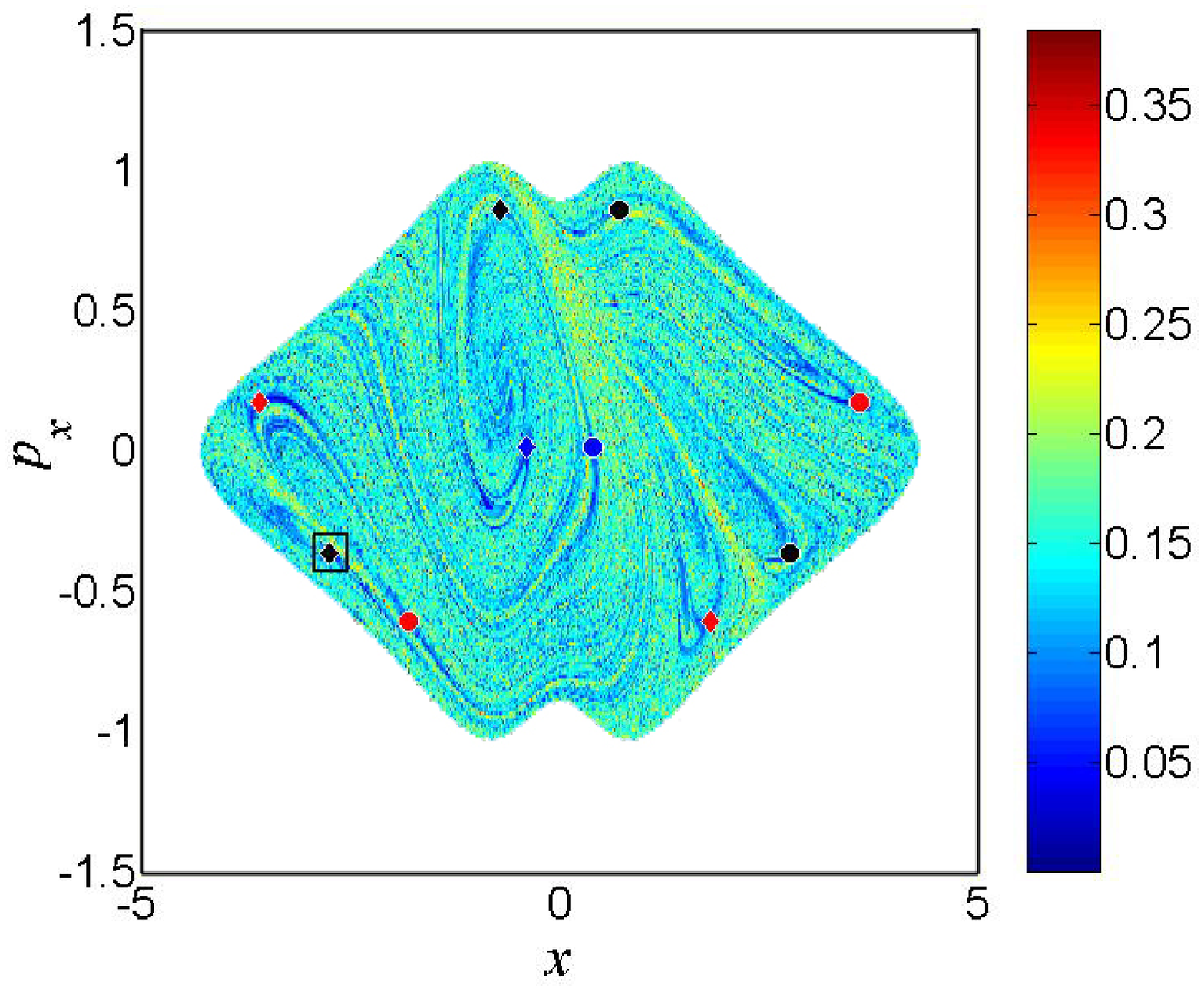}
		\includegraphics[width = \linewidth]{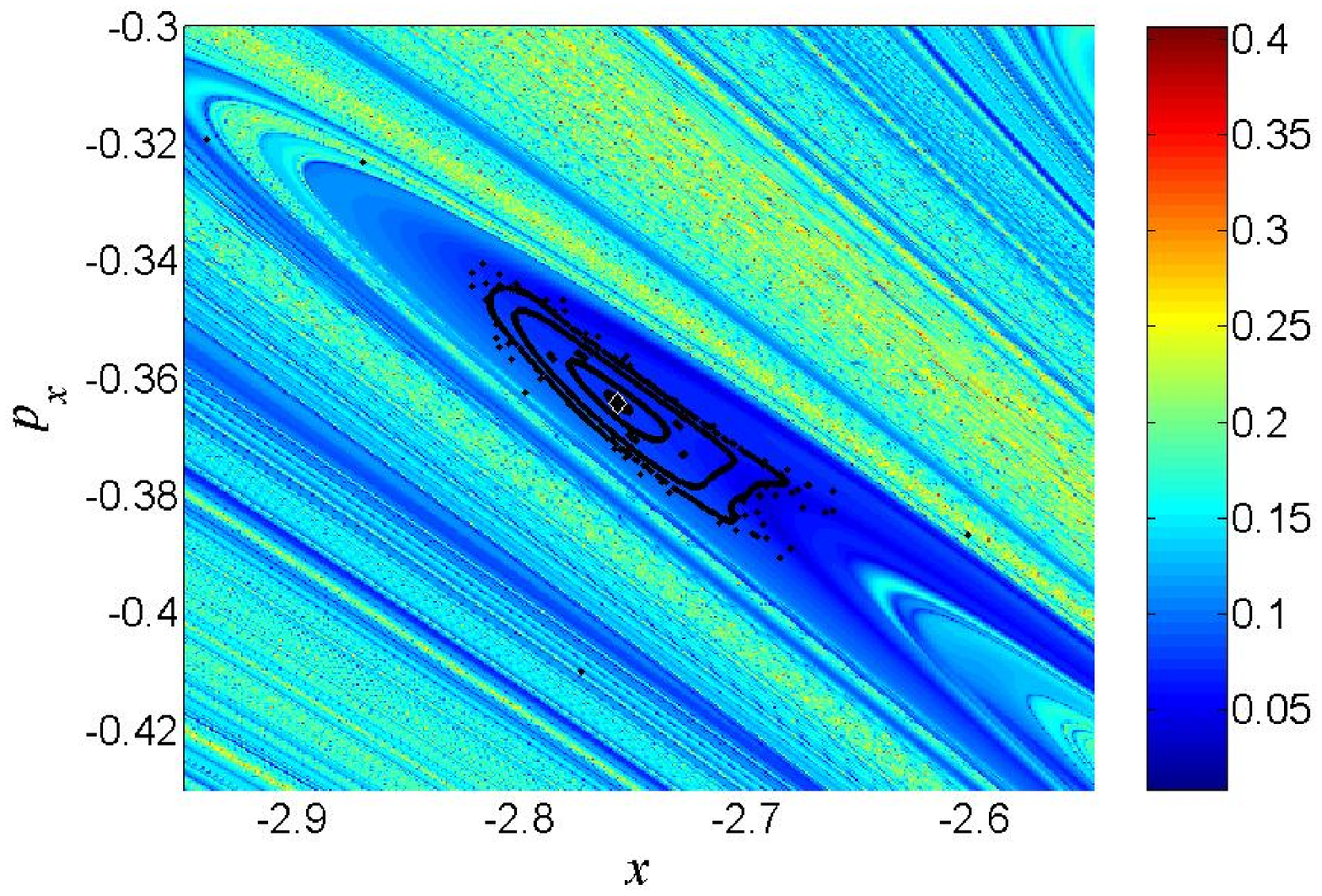}
	\caption{\label{fig:Fig3}
	Upper panel~: FTL map of Hamiltonian~(\ref{Eq:Hamiltonian}) without the field at time~$t=50$~a.u. in the plane~$(x, p_{x})$ with~$y=0$. The full circles indicate (with the same color code) the intersections of the periodic orbits of Fig.~\ref{fig:PO} with the Poincar\'e section at $y=0$~: The full diamonds indicate the symmetrical orbits from the one of Fig.~\ref{fig:PO} with the symmetry $(x,y,p_x,p_y)\mapsto (-x,-y,-p_x,-p_y)$. Lower panel~: Enlargement of the FTL map around a region indicated by a square in the upper panel, and Poincar\'e sections of some regular trajectories in the elliptic island. }
\end{figure}

\begin{figure}
	\centering
		\includegraphics[width = \linewidth]{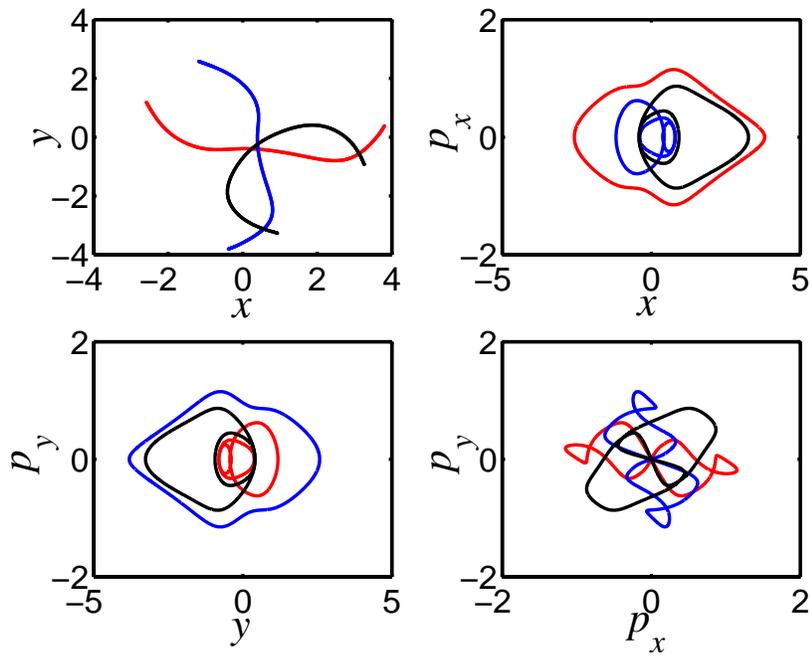}
	\caption{\label{fig:PO}
Projections of the periodic orbits $O_e$ (black curves), $O_{x,1}$ (red curves) and $O_{y,1}$ (blue curves) in the $(x,y)$ plane (upper left panel), $(x,p_x)$ plane (upper right panel), $(y,p_y)$ plane (lower left panel), and $(p_x,p_y)$ plane (lower right panel).}
\end{figure}

\begin{figure}
	\centering
		\includegraphics[width = \linewidth]{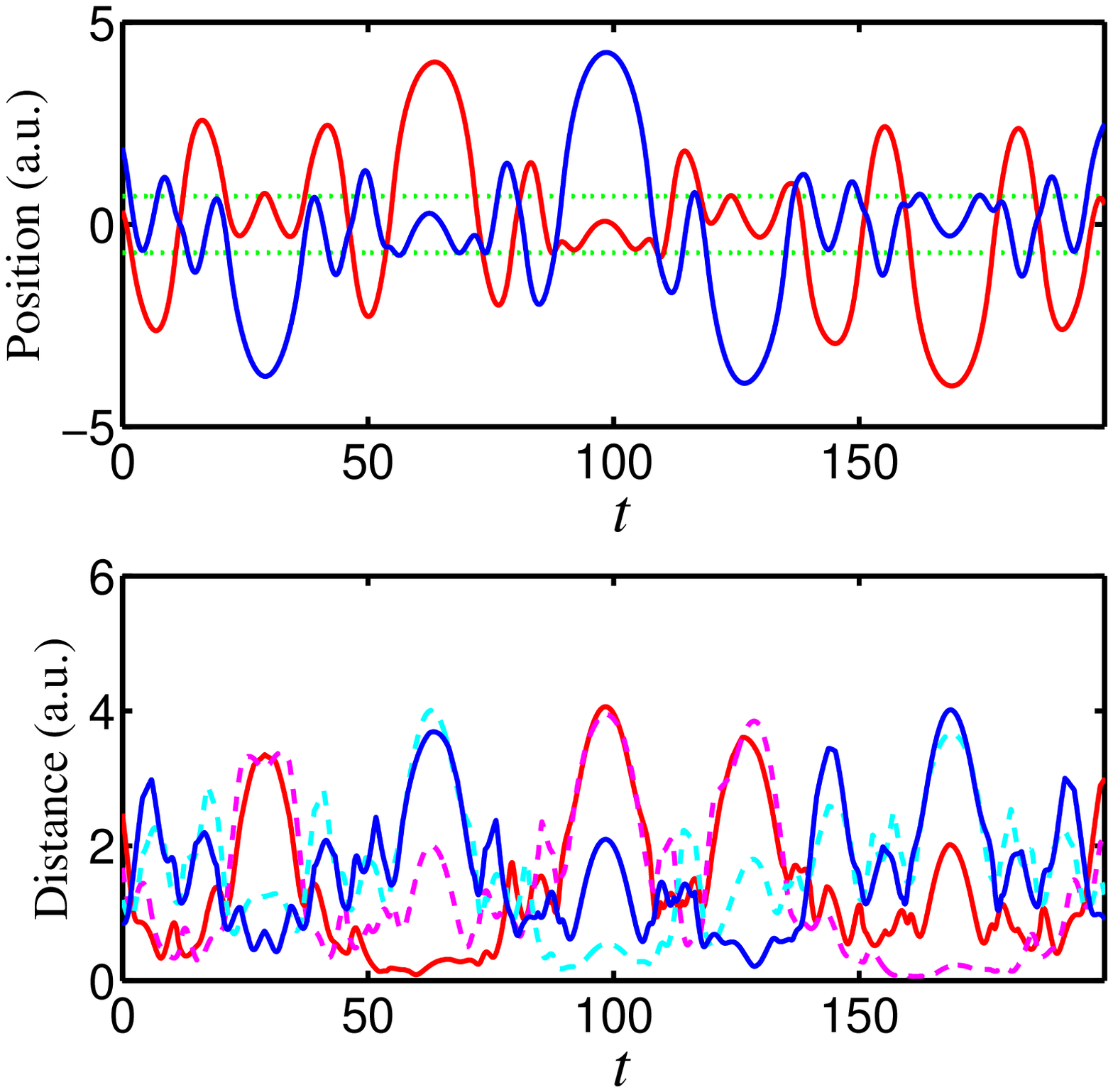}
		\caption{\label{fig:distPO} Upper panel~: Positions $x$ (red online) and $y$ (blue online) of the two electrons as a function of time of a typical trajectory of Hamiltonian~(\ref{Eq:Hamiltonian}) with $d=1$ and without the field. Lower panel~: Distance in phase space of the trajectory depicted in the upper panel versus time to the four periodic orbits~$O_{x,1}$, $O_{x,2}$, $O_{y,1}$, $O_{y,2}$. For each distance, the color code and line style follow the one in Fig.~\ref{fig:PO}, the dashed curves (magenta and cyan) correspond to the distance to the symmetrical periodic orbits $O_{x,2}$ and $O_{y,2}$ respectively. The horizontal dashed green lines indicate the positions of the nuclei. Time is in atomic units.
	}
\end{figure}

We first analyze the dynamics of Hamiltonian~(\ref{Eq:Hamiltonian}) without the field~($E_{0} = 0$). This Hamiltonian system has two degrees of freedom, and therefore lends itself very well to an analysis using Poincar\'e sections. Here we use a different tool to analyze the dynamics, by considering the linear stability properties such as obtained by the finite-time Lyapunov (FTL) exponents~\cite{chaosbook,Shch06}. 
These FTL exponents are obtained by integrating the tangent flow together with the equations of motion for ${\bf X}=(x,y,p_x,p_y)$~:
\begin{eqnarray}
   \dot{\bf X} & = & {\bf F}({\bf X},t),          \label{Eq_motion} \\
   \dot{J} & = & DF({\bf X},t)  J, \label{Tangent_flow}
\end{eqnarray}
where Eq.~(\ref{Eq_motion}) are the equations of motion, and Eq.~(\ref{Tangent_flow}) is the tangent flow where $DF({\bf X},t)$ is the matrix of variations of the generalized velocity field ${\bf F}$ at the point~$\bf X$ and time $t$, i.e.\ $DF_{ij}=\partial F_i/\partial X_j$. The initial condition for the integration of the tangent flow is $J_{0} = \mathbb{I}_{4}$, the four dimensional identity matrix. The (maximum) FTL exponent at time $t$ for the initial conditions ${\bf X}_0$ is equal to $l(t;{\bf X}_0) = \log|\lambda(t;{\bf X}_0)|/t$ where~$\lambda(t;{\bf X}_0)$ is the eigenvalue of the Jacobian matrix $J$ at time~$t$ with the largest norm. The way to analyze the dynamics using these exponents is to represent maps of FTL exponents as functions of the initial conditions ${\bf X}_0$ at a fixed time $t$. These maps (called FTL maps) quantify the (linear) instability of some regions and highlight invariant objects. 
A typical FTL map is depicted in Fig.~\ref{fig:Fig3} for Hamiltonian~(\ref{Eq:Hamiltonian}) without the field. It clearly displays strong and global chaos by showing fine details of the stretching and folding of trajectories~\cite{chaosbook}. A closer inspection shows some regions associated with a very small FTL exponents (blue regions and lower panel of Fig.~\ref{fig:Fig3}). A Poincar\'e section reveals the presence of an elliptic island which surrounds an elliptic periodic orbit $O_e$, represented on Fig.~\ref{fig:PO} with a black curve. However such regions are too small to play a major role in the dynamics of a typical trajectory. In fact, the motion without the field is guided by four weakly hyperbolic periodic orbits. These four important periodic orbits are denoted~$O_{x,1}$, $O_{x,2}$, $O_{y,1}$ and~$O_{y,2}$, and the projections of $O_{x,1}$ and $O_{y,1}$ are displayed in Fig.~\ref{fig:PO}. These four periodic orbits are obtained from one single periodic orbit through the symmetries of the equations of motion:~$(x,y,p_{x},p_{y}) \mapsto (y,x,p_{y},p_{x})$ and $(x,y,p_{x},p_{y}) \mapsto (-x,-y,-p_{x},-p_{y})$. It also means that the representation of these orbits on the plane~$(y,p_{y})$ can be deduced from the one on the plane~$(x,p_{x})$ by inverting the coordinates~$x$ and~$y$. The two periodic orbits~$O_{x,1}$ and $O_{x,2}$ (stretched in the $x$-direction) have an outer projection in the $(x,p_x)$ plane and an inner one in the $(y,p_y)$ plane (see the red curves in Fig.~\ref{fig:PO}). By symmetry, the same holds for $O_{y,1}$ and $O_{y,2}$, swapping the role of $x$ and $y$.  Figure~\ref{fig:distPO} represents the distance of a typical trajectory with the four periodic orbits mentioned above. It shows that at each time, the trajectory is close to one of these orbits with rapid transitions between them. This means that the motion on each of these periodic orbits, and consequently of a typical trajectory, is composed of one electron moving close to the nucleus (the inner electron) and the other one further moving away (the outer electron), with quick exchanges of the roles of each electron.

The reason why these hyperbolic periodic orbits are important is that they are short and weakly hyperbolic periodic orbits (their Greene residue~\cite{gree79} is equal to $1.1$). Hence a typical trajectory passing nearby one of these orbits mimics (by continuity) the motion on the periodic orbit and then slowly escapes (depending on its escape rate and hence on its stability properties) from this orbit following its unstable manifold.

\paragraph*{Single ionization}
By switching on the field, the outer electron is picked up and swept away from the nucleus. Consequently, its effective Hamiltonian is composed of its kinetic energy and the interaction potential of the laser field~:
\begin{equation} \label{Eq:H1}
   \mathcal{H}_{1} = \frac{p_{x}^{2}}{2} + E_{0} x f(t) \sin \omega t
\end{equation}
We notice that the trajectories of Hamiltonian~$\mathcal{H}_{1}$ can be explicitly computed. They are composed of a linear escape from the nuclei~(at time~$t_{0}$) modulated by the action of the field~\cite{Cork93} (see trajectories after collision in Fig.~\ref{fig:Fig2}).

\begin{figure}
	\centering
	 \includegraphics[width = \linewidth]{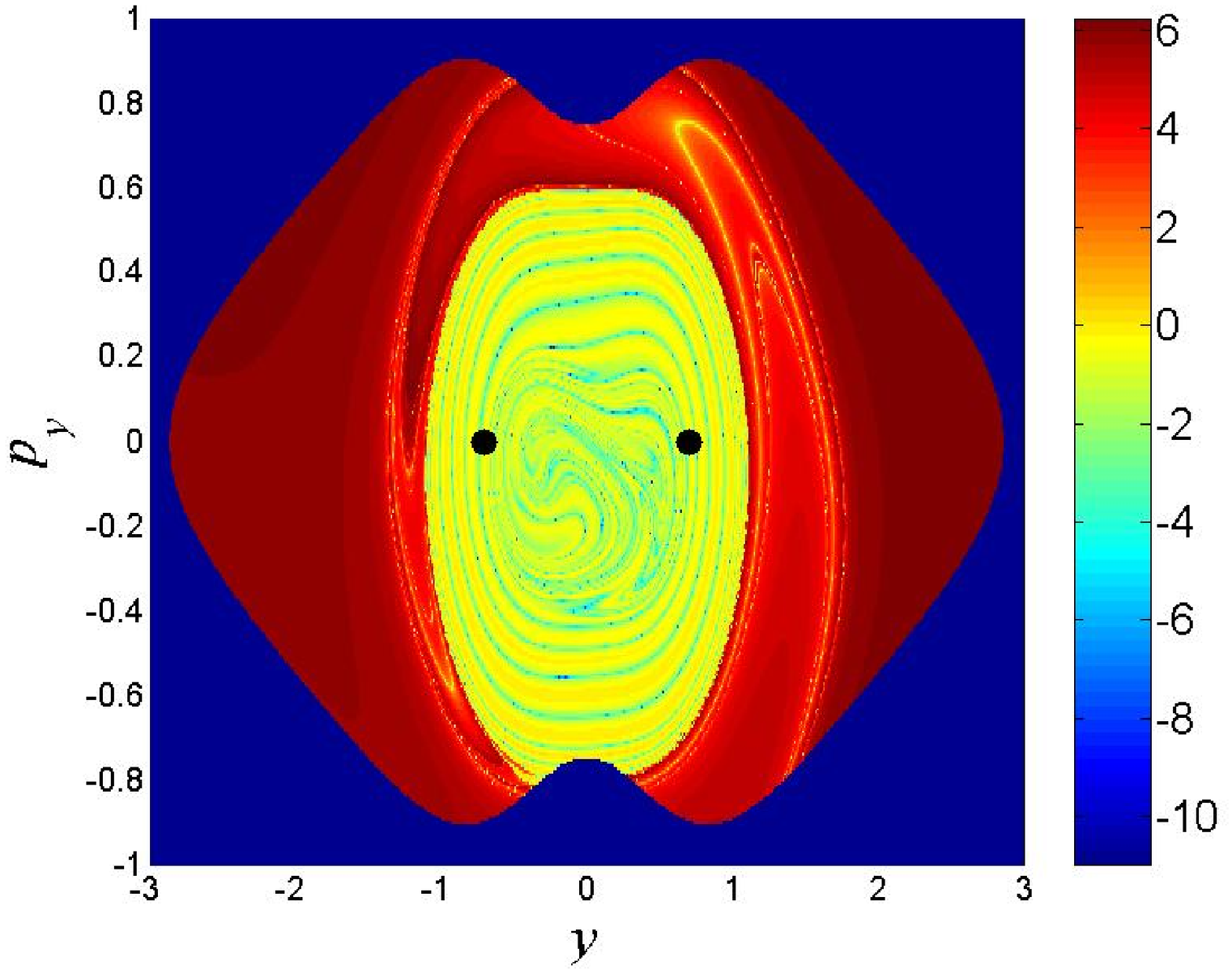}
		\includegraphics[width = \linewidth]{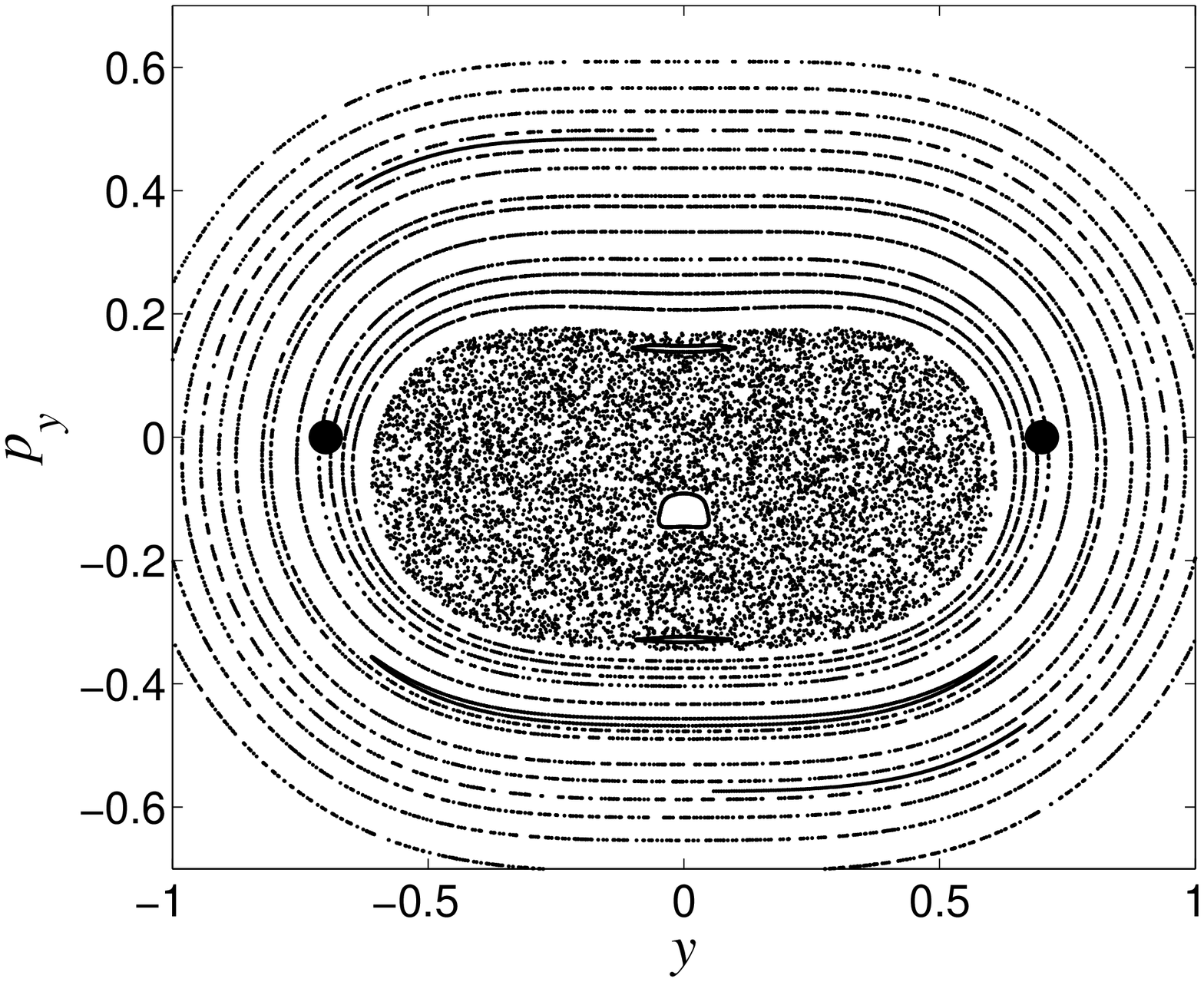}
	\caption{\label{fig:Fig4}
	Upper panel~: Contour plot of the electron location~$y(t)$ after two laser cycles in the plane of initial conditions $(y_0,p_{y,0})$ of Hamiltonian~(\ref{Eq:H2}) for $I = 3 \times 10^{15}\ \mbox{W} \cdot \mbox{cm}^{-2}$ and $\omega=0.0584$. The color code is on a logarithmic scale. Lower panel~: Poincar\'e section~(stroboscopic plot with a period of one laser cycle) of some trajectories for the same Hamiltonian as the upper panel. The black dots indicate the positions of the nuclei.}
\end{figure}

\begin{figure}
	\centering
		\includegraphics[width = \linewidth]{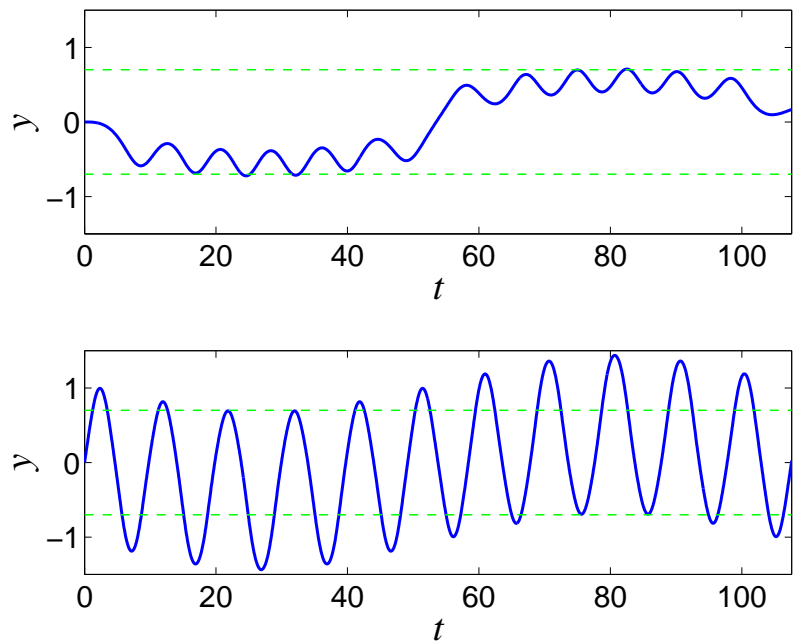}
	\caption{\label{fig:typ}
	Two types of trajectories of Hamiltonian~(\ref{Eq:H2}) for~$I = 10^{15}\ \mbox{W} \cdot \mbox{cm}^{-2}$ and $\omega=0.0584$ during a laser cycle~: One inside the inner chaotic region (upper panel), and one on an outer invariant torus (lower panel). The horizontal dashed lines (green online) indicate the positions of the two nuclei.}
\end{figure}

\begin{figure}
	\centering
		\includegraphics[width = \linewidth]{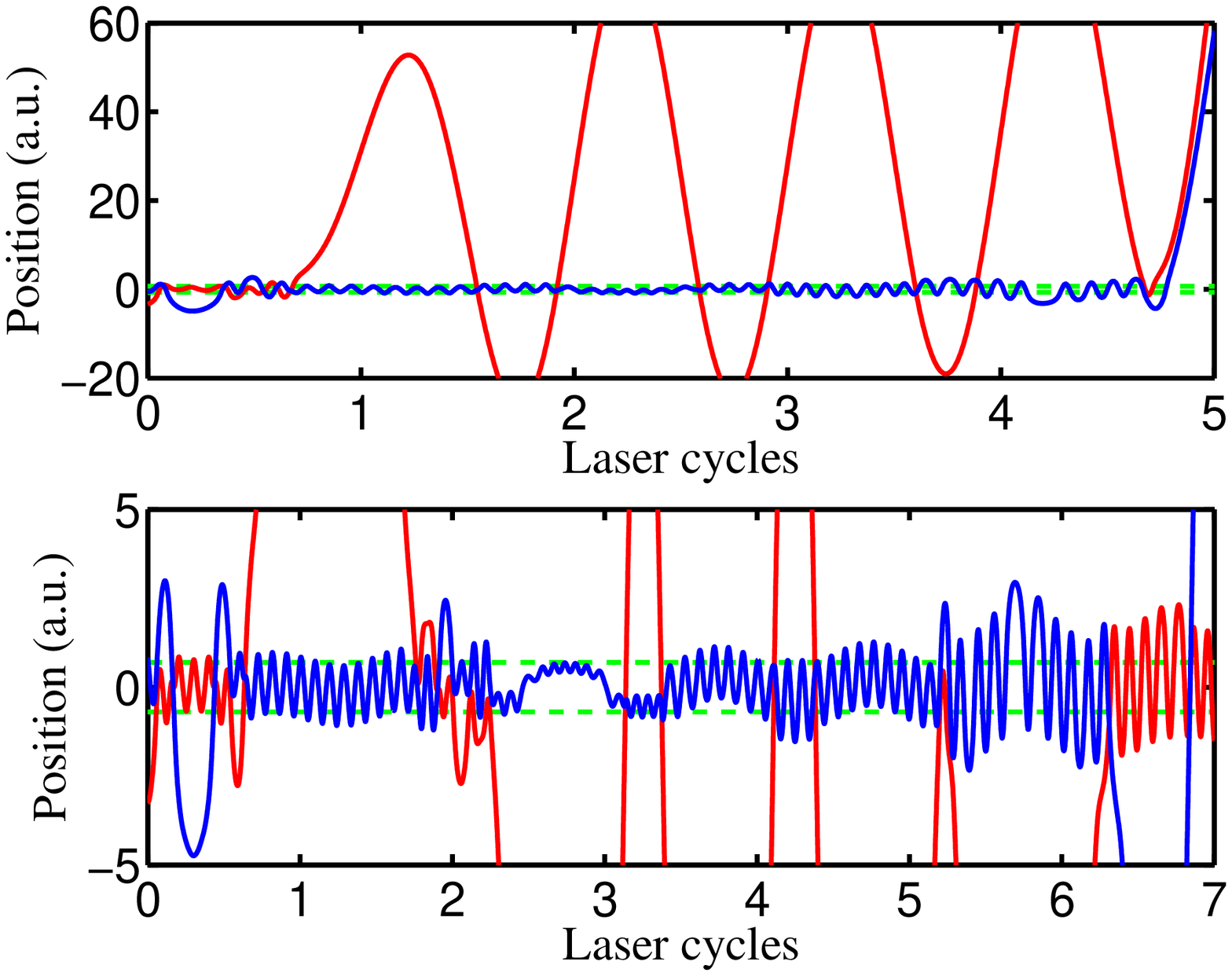}
	\caption{\label{fig:Fig5}
	Two typical trajectories of Hamiltonian~(\ref{Eq:Hamiltonian}) for~$I = 10^{15}\ \mbox{W} \cdot \mbox{cm}^{-2}$ and $\omega=0.0584$ for initial conditions in the ground state energy of the $H_{2}$ molecule. The horizontal dashed lines (green online) indicate the positions of the two nuclei.}
\end{figure}

\paragraph*{Sequential double ionization~(SDI)}
Since the outer electron is far from the core region, the effective Hamiltonian for the inner electron contains the potential of the nuclei and the interaction with the laser field~:
\begin{eqnarray} \label{Eq:H2}
   \mathcal{H}_{2} & = & \frac{p_{y}^{2}}{2} + y E_{0} \sin \omega t  \nonumber \\
                   &   & - \frac{1}{\sqrt{ \left(y-R/2 \right)^{2} + 1}}
                         - \frac{1}{\sqrt{ \left(y+R/2 \right)^{2} + 1}}
\end{eqnarray}
In the absence of the field~($I = 0$),~$\mathcal{H}_{2}$ is also integrable and the inner electron is confined on a periodic orbit, for almost all the initial conditions. By switching on the field, the dynamics associated with $\mathcal{H}_2$ is significantly affected. A contour plot of $y$ after two laser cycles and a Poincar\'e section are represented on Fig.~\ref{fig:Fig4} for $I=3\times 10^{15}\ \mbox{W} \cdot \mbox{cm}^{-2}$. We notice the following features~: First, most of the regular trajectories evolve on invariant tori. Second, most of the inner tori are broken by the field, creating a bounded chaotic region between the two nuclei. These two types of trajectories (regular and bounded chaotic) do not ionize. Third, the outermost invariant tori are broken since the motion is unbounded sufficiently far away from the nuclei. These are the ionizing trajectories. During a full laser cycle, we expect two types of non-ionizing trajectories represented on Fig.~\ref{fig:typ}~: A typical motion in between the nuclei (upper panel) and another surrounding the two nuclei (lower panel). This is what is observed on typical trajectories like the one represented on the lower panel of Fig.~\ref{fig:Fig5}. 

\begin{figure}
	\centering
		\includegraphics[width = \linewidth]{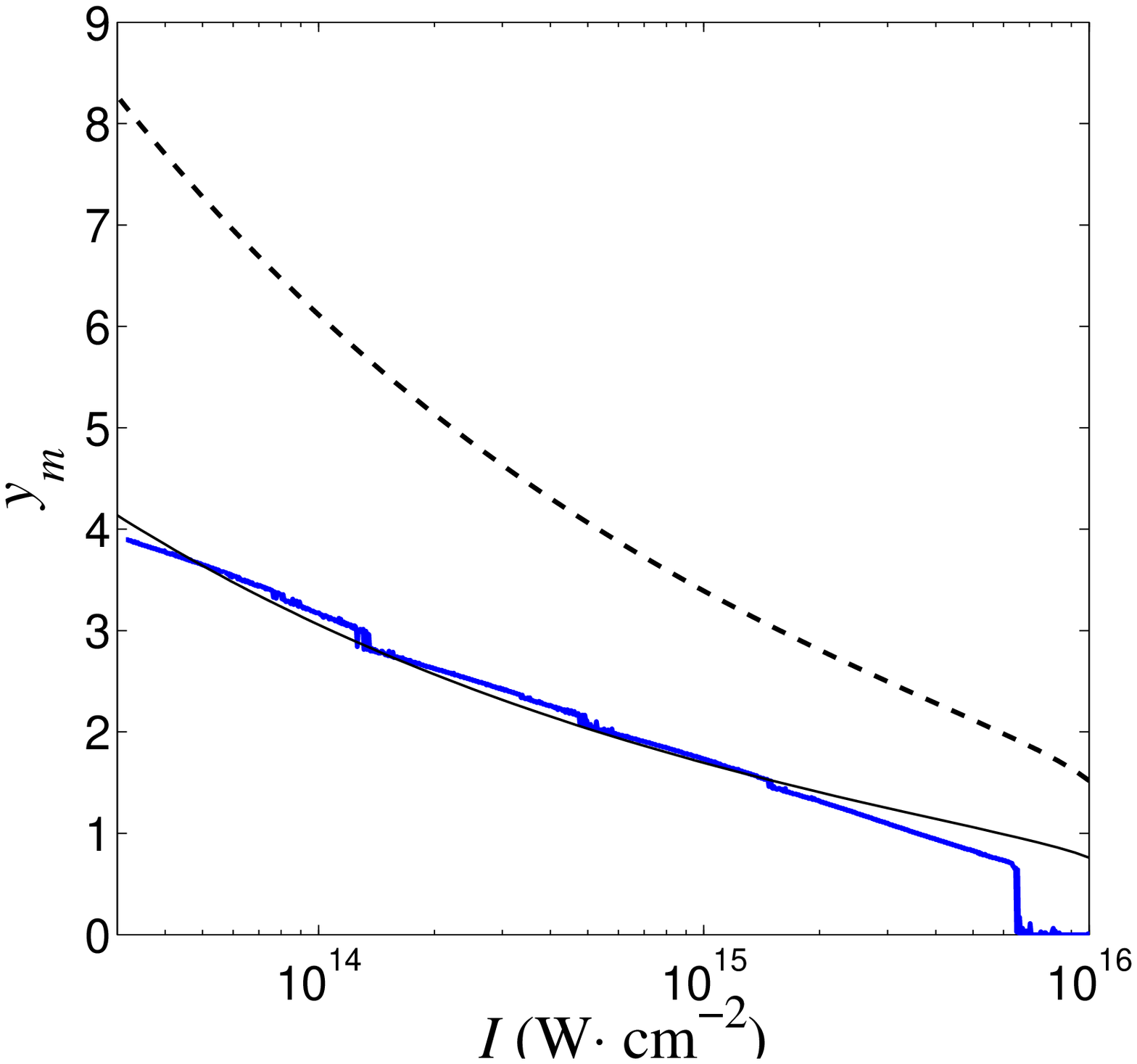}
	\caption{\label{fig:Fig_ym}
	Numerical estimates of $y_m$ as a function of the laser intensity $I$ for Hamiltonian~(\ref{Eq:Hamiltonian}) for $\omega=0.0584$ (continuous blue curve). The qualitative approximation of $y_m$ given by Eq.~(\ref{Eq:YmAndE0}) (dashed curve) and the same qualitative approximation divided by two (thin continuous curve). }
\end{figure}

By varying the intensity, the picture of the phase space of Hamiltonian~(\ref{Eq:H2}) evolves in the following way~: If the laser intensity~$I$ is too small, then the phase space is surrounded by invariant tori~(with a bounded chaotic region between the two nuclei) and no sequential double ionization can occur. The sequential double ionization probability depends the on the size of the regular region, and hence on~$I$. For large values of the intensity, the regular zone vanishes and after some critical intensity, after which there are no invariant tori bounding the motion of the inner electron. Consequently, only sequential double ionization is observed in this high intensity regime.
More specifically, the size of the region where the trajectories are bounded is quantified by $y_m$ (defined as the maximum position $y$ of the outermost invariant torus of Hamiltonian~(\ref{Eq:H2}) on the Poincar\'e section). A numerical approximation to $y_m$ is obtained by integrating trajectories at $p_y=0$ (since the domain is approximately symmetric with respect to $p_y=0$) and monitoring the ones which do not keep a bounded value for the position after the duration of the laser pulse. This numerical estimate of $y_m$ versus the intensity $I$ is represented on Fig.~\ref{fig:Fig_ym} (continuous curve). 

\begin{figure}
	\centering
	\setlength{\unitlength}{1mm}
        \begin{picture}(80,61.5)
        \put(0,0){\includegraphics[width=80mm]{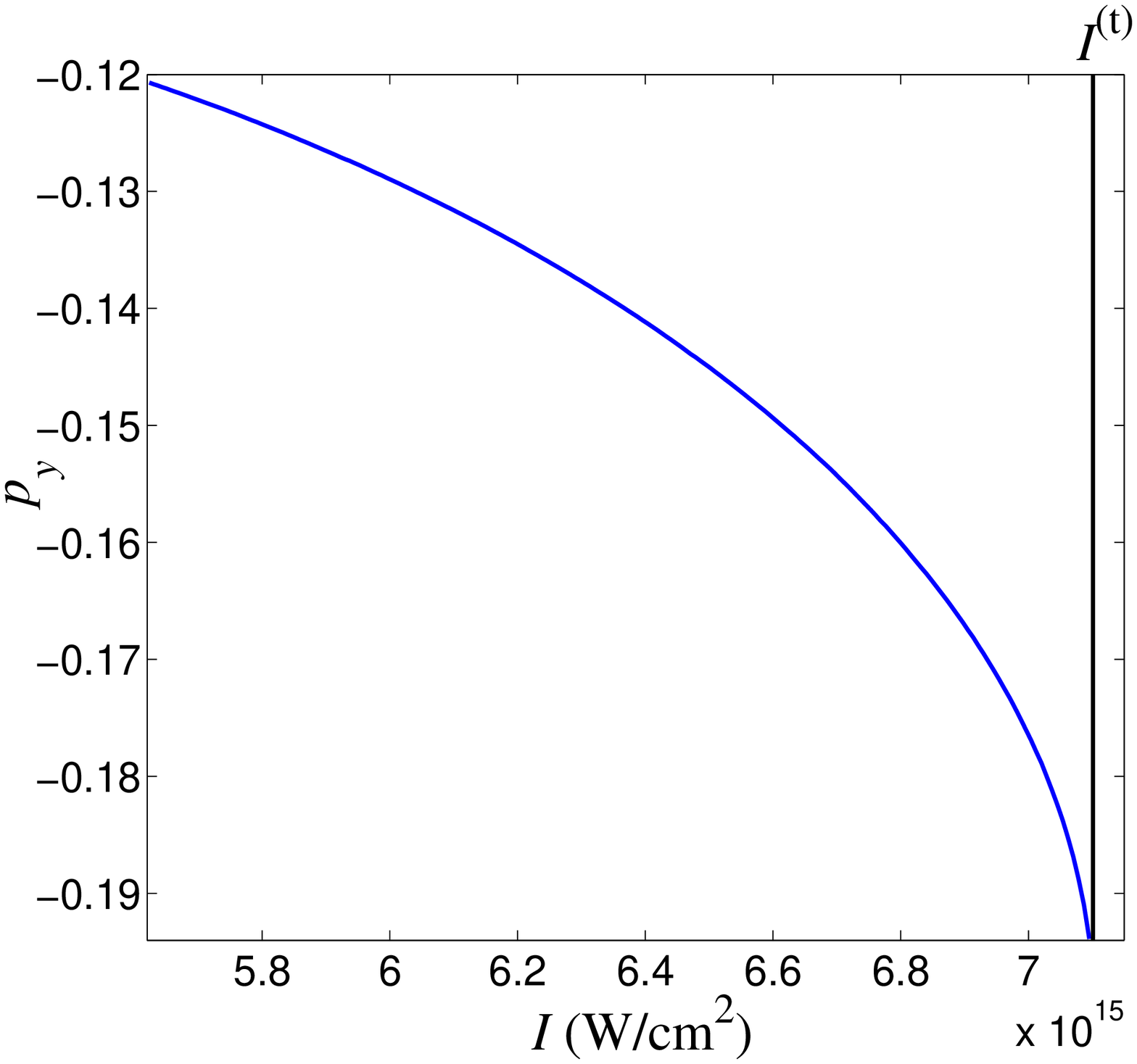}}
        \put(11,9){\includegraphics[width=35mm]{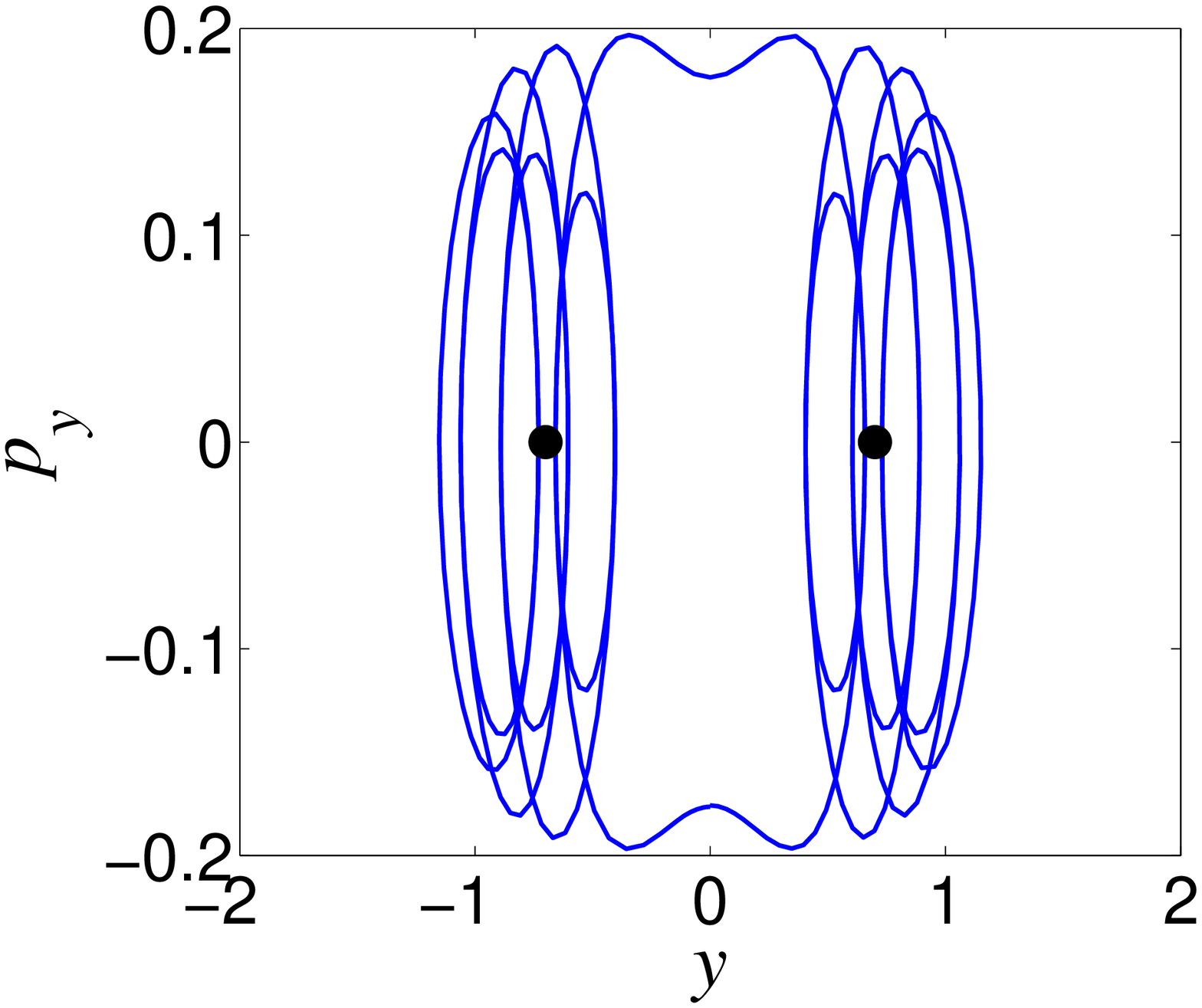}}
        \end{picture}
	\caption{\label{fig:Fig7}
	 Momentum of the central periodic orbit (on the Poincar\'e section) of Hamiltonian~(\ref{Eq:H2}) as a function of the laser intensity. The vertical line on the lower panel indicates the intensity~$I^{(t)} = 7.10 \times 10^{15}\ \mbox{W} \cdot \mbox{cm}^{-2}$ such that for~$I \geq I^{(t)}$, complete unhindered SDI is expected. The black dots indicate the positions of the two nuclei. Inset~: Projection of the central periodic orbit at~$I = 7 \times 10^{15}\ \mbox{W} \cdot \mbox{cm}^{-2}$ in the~$(y, p_{y})$-plane.}
\end{figure}

A rough approximation to~$y_{m} = y_{m} \left( E_{0} \right)$ is given implicitly by the value where the potential of Hamiltonian~(\ref{Eq:H2}) is locally maximum, i.e.~:
\begin{equation} \label{Eq:YmAndE0}
   E_ {0} = \frac{y_{m} - R/2}{\left( \left(y_{m} -R/2 \right)^{2} + 1 \right)^{3/2}}
          + \frac{y_{m} + R/2}{\left( \left(y_{m} +R/2 \right)^{2} + 1 \right)^{3/2}},
\end{equation}
independently of the laser frequency. The solution of the above equation is represented as a function of the laser intensity $I$ in Fig.~\ref{fig:Fig_ym} (dashed curve). The observation is that the theoretical prediction captures the right order of magnitude, but for $\omega=0.0584$, this approximation is quantitatively not accurate since the estimates are about twice the actual values. Of course the quality of the approximation depends on the chosen frequency $\omega$. As the frequency $\omega$ is decreased toward zero, the approximation becomes better. \\
Equation~(\ref{Eq:YmAndE0}) has a solution for $I\leq 1.09\times 10^{16}\ \mbox{W} \cdot \mbox{cm}^{-2}$. For higher intensities, we expect complete double ionization. This is the case of Fig.~\ref{fig:Fig1}. However, this is an upper bound since complete double ionization appears to happen at smaller intensities. This emerges from the numerical computation of $y_m$ where the size of the regular zone drops to zero for intensities larger than $7.10 \times 10^{15}\ \mbox{W} \cdot \mbox{cm}^{-2}$. The origin of this phenomenon is explained by looking at the central periodic orbit which organizes the bounded motion. This periodic orbit has the same period as the field (and therefore intersects the Poincar\'e section at only one point). It is located at $y=0$. In order to determine its momentum, we use a Newton-Raphson method for the determination of periodic orbits~\cite{chaosbook}. The results are represented on Fig.~\ref{fig:Fig7}. We see that for a wide range of intensities the momentum $p_y$ of the periodic (on the Poincar\'e section) does not change significantly. Around $I=7\times 10^{15}\ \mbox{W} \cdot \mbox{cm}^{-2}$ it starts decreasing and drops to $-\infty$ at $I^{(t)}=7.10 \times 10^{15}\ \mbox{W} \cdot \mbox{cm}^{-2}$. At this value, complete sequential double ionization occurs, in good agreement with Fig.~\ref{fig:Fig1}. 

\paragraph*{Nonsequential double ionization~(NSDI)}
As noted before, when the field is turned on, its action is concentrated on only one electron, the outer one, as a first step. The field drives the outer electron away from the nuclei leaving the inner electron nearly unaffected by the field (or at least bounded if it is between the two nuclei) because its position remains small. From the recollision process~\cite{Cork93,Baie06}, the outer electron might come back close to the nuclei during the pulse plateau. In this case, it transfers a part of its energy to the inner electron through the electron-electron interaction term (a recollision). From then on, two outcomes are generically possible~: If the energy brought in by the outer electron is sufficient for the other electron to escape from the inner region (as in the lower panel of Fig.~\ref{fig:Fig2}), then it might ionize together with the outer electron. The maximum energy~$\mathcal{E}_{x}$ of the outer electron when it returns to the inner region is obtained from Hamiltonian~(\ref{Eq:H1}) and is~$\mathcal{E}_{x} = \kappa U_{p}$ where~$U_{p} = E_{0}^{2}/(4 \omega^{2})$ is the ponderomotive energy and~$\kappa = 3.17314 \ldots$~\cite{Cork93,Band05}. We complement the recollision scenario (which focuses on the outer electron) by providing the phase space picture of the inner electron as accurately described by Hamiltonian~(\ref{Eq:H2})~: In order to ionize the inner electron in the most efficient way, the energy brought back by the outer electron has to be of order of the energy between the center ($y=0$) and the boundary of the bounded inner region~($y = y_{m}$) of the phase space of Hamiltonian~$\mathcal{H}_{2}$~(see Fig.~\ref{fig:Fig4}). This energy difference is equal to~:
\begin{eqnarray} \label{Eq:EnerDiff}
   \Delta \mathcal{E}_{y} & = & \frac{2}{\sqrt{R^{2}/4+1}}
                            - \frac{1}{\sqrt{\left( y_{m} - R/2 \right)^{2}+1}} \nonumber \\
                          & - & \frac{1}{\sqrt{\left( y_{m} + R/2 \right)^{2}+1}}.
\end{eqnarray}
The equal-sharing relation which links the classical picture of the outer electron~$x$ with the one of the inner electron~$y$,
\begin{equation} \label{Eq:EqualSharing}
   \Delta \mathcal{E}_{y} = \frac{\mathcal{E}_{x}}{2} = \kappa \frac{E_{0}^{2}}{8 \omega^{2}}
\end{equation}
defines~(through an implicit equation) the expected value of the field~$E_{0}^{(c)}$ and hence the corresponding intensity $I^{(c)}$ for maximal NSDI because it describes the case when each outer electron brings back enough energy to ionize the innermost electrons, while keeping enough energy to potentially remain ionized itself. However, fulfilling this energy requirement does not guarantee NSDI~: The outcome depends on the number and the efficiency of recollisions.  In order to estimate $I^{(c)}$, we first use the qualitative approximation of $y_m$ given by Eq.~(\ref{Eq:YmAndE0}). For $\omega=0.0584$, this intensity is equal to $3.60\times 10^{14}\ \mbox{W} \cdot \mbox{cm}^{-2}$ which is in good agreement with the maximum of the knee in Fig.~\ref{fig:Fig1} (green dashed curve). Now we consider the numerically determined values for $y_m$ (see Fig.~\ref{fig:Fig_ym}). By solving Eq.~(\ref{Eq:EqualSharing}), the intensity $I^{(c)}$ is found at $2.60\times 10^{14}\ \mbox{W} \cdot \mbox{cm}^{-2}$ (green continuous curve in Fig.~\ref{fig:Fig1}) in qualitative agreement with the above value obtained using the qualitative determination of $y_m$.

Next we investigate how this critical intensity $I^{(c)}$ varies with the laser frequency and highlight how this intensity varies with the distance $R$ between the two nuclei. For that purpose, we expand the equal sharing relation~(\ref{Eq:EqualSharing}) using to small parameters: $R^2/8$ and $\eta_0=2\omega/\sqrt{\kappa}$ (which are assumed to be of the same order). The leading order of Eq.~(\ref{Eq:EnerDiff}) is
$$
\Delta {\mathcal E}_y\approx 2\left(1-\frac{R^2}{8}\right)-\frac{2}{\sqrt{y_m^2+1}},
$$
and the corresponding expansion for the approximation $y_m=y_m(E_0)$ given by Eq.~(\ref{Eq:YmAndE0}) is
$$
E_0=\frac{2y_m}{(y_m^2+1)^{3/2}}.
$$ 

If we denote $\eta=1/\sqrt{y_m^2+1}$, then Eq.~(\ref{Eq:EqualSharing}) is expanded into
$$
\eta^4(1-\eta^2)=\eta_0^4\left(1-\eta-\frac{R^2}{8}\right).
$$
An expansion of a solution of the above equation is given by
$$
\eta=\eta_0-\frac{\eta_0}{4}\left(\eta_0+\frac{R^2}{8}\right),
$$
which translates into an expansion for the critical amplitude 
\begin{equation}
\label{Eq:approxEc}
E_0^{(c)}\approx \frac{4\omega}{\sqrt{\kappa}}-\left(\frac{2\omega}{\sqrt{\kappa}}\right)^{3/2}-\frac{2\omega}{\sqrt{\kappa}}\frac{R^2}{8}.
\end{equation}
We notice that the expansion of $E_0^{(c)}$ at $R=0$ is the same as the one found for the helium atom~\cite{maug09} since the Hamiltonian model for the helium atom is the $R=0$ limit of Hamiltonian~(\ref{Eq:Hamiltonian}) .
As the distance $R$ between the two nuclei increases, the location of the knee is displaced toward lower intensities, and this displacement is proportional to $R^2$. This is actually what is observed when we compared the values of $I^{(c)}$ for the helium atom~\cite{maug09} ($I^{(c)}(R=0)=4.60 \times 10^{14}\ \mbox{W} \cdot \mbox{cm}^{-2}$ for $\omega=0.0584$) and for the ${\rm H}_2$ molecule.
For $\omega=0.0584$, the approximate value of $I^{(c)}$ given by Eq.~(\ref{Eq:approxEc}) for $R=1.4$ is $ 3.40 \times 10^{14}\ \mbox{W} \cdot \mbox{cm}^{-2}$ which is in good agreement the numerical solution of Eq.~(\ref{Eq:EqualSharing}).

\section{Conclusion}
The classical picture for the sequential and nonsequential double ionization of ${\rm H}_2$ is obtained by complementing the recollision scenario for the outer electron with the phase space picture of the inner electron. By finding the organizing principles of the classical dynamics, we arrived at two predictions for the characteristic intensities of the knee-shaped double ionization probability curve~: the intensity after which the complete double ionization is expected, and the intensity where the non-sequential double ionization is predicted to be maximum. Very good agreement is found by comparing our predictions with the direct integration of a large assembly of trajectories. It should be noted that this scenario follows closely the one of the helium atom, implying the existence of a general mechanism for nonsequential double ionization of atoms and molecules. 

Based on these findings, we believe that advanced methods of classical mechanics and the diagnostics we use here are worth pursuing further  because classical mechanics, with its advantageous scaling with system size, may well become a useful tool for understanding some aspects of molecular systems too complex for a full quantal treatment with contemporary computing resources.

\section*{Acknowledgments}
CC acknowledges financial support from the PICS program of the CNRS. This work is partially funded by NSF.



\begin{thebibliography}{47}
\expandafter\ifx\csname natexlab\endcsname\relax\def\natexlab#1{#1}\fi
\expandafter\ifx\csname bibnamefont\endcsname\relax
  \def\bibnamefont#1{#1}\fi
\expandafter\ifx\csname bibfnamefont\endcsname\relax
  \def\bibfnamefont#1{#1}\fi
\expandafter\ifx\csname citenamefont\endcsname\relax
  \def\citenamefont#1{#1}\fi
\expandafter\ifx\csname url\endcsname\relax
  \def\url#1{\texttt{#1}}\fi
\expandafter\ifx\csname urlprefix\endcsname\relax\def\urlprefix{URL }\fi
\providecommand{\bibinfo}[2]{#2}
\providecommand{\eprint}[2][]{\url{#2}}

\bibitem[{\citenamefont{Posthumus}(2004)}]{post04}
\bibinfo{author}{\bibfnamefont{J.~H.} \bibnamefont{Posthumus}},
  \bibinfo{journal}{Rep. Prog. Phys.} \textbf{\bibinfo{volume}{67}},
  \bibinfo{pages}{623} (\bibinfo{year}{2004}).

\bibitem[{\citenamefont{Bandrauk and Corkum}(2006)}]{band06u}
\bibinfo{author}{\bibfnamefont{A.~D.} \bibnamefont{Bandrauk}} \bibnamefont{and}
  \bibinfo{author}{\bibfnamefont{P.~B.} \bibnamefont{Corkum}}
  (\bibinfo{year}{2006}), \bibinfo{note}{unpublished}.

\bibitem[{\citenamefont{Kling and Vrakking}(2008)}]{klin08}
\bibinfo{author}{\bibfnamefont{M.~F.} \bibnamefont{Kling}} \bibnamefont{and}
  \bibinfo{author}{\bibfnamefont{M.~J.~J.} \bibnamefont{Vrakking}},
  \bibinfo{journal}{Ann. Rev. Phys. Chem.} \textbf{\bibinfo{volume}{59}},
  \bibinfo{pages}{463} (\bibinfo{year}{2008}).

\bibitem[{\citenamefont{Bandrauk et~al.}(2004)\citenamefont{Bandrauk,
  Chelkowski, and Nguyen}}]{band04}
\bibinfo{author}{\bibfnamefont{A.~D.} \bibnamefont{Bandrauk}},
  \bibinfo{author}{\bibfnamefont{S.}~\bibnamefont{Chelkowski}},
  \bibnamefont{and} \bibinfo{author}{\bibfnamefont{H.~S.}
  \bibnamefont{Nguyen}}, \bibinfo{journal}{Int. J. Quant. Chem.}
  \textbf{\bibinfo{volume}{100}}, \bibinfo{pages}{834} (\bibinfo{year}{2004}).

\bibitem[{\citenamefont{Chelkowski et~al.}(2006)\citenamefont{Chelkowski,
  Yudin, and Bandrauk}}]{chel06}
\bibinfo{author}{\bibfnamefont{S.}~\bibnamefont{Chelkowski}},
  \bibinfo{author}{\bibfnamefont{G.~L.} \bibnamefont{Yudin}}, \bibnamefont{and}
  \bibinfo{author}{\bibfnamefont{A.~D.} \bibnamefont{Bandrauk}},
  \bibinfo{journal}{J. Phys. B: At. Mol. Opt. Phys.}
  \textbf{\bibinfo{volume}{39}}, \bibinfo{pages}{S409} (\bibinfo{year}{2006}).

\bibitem[{\citenamefont{Kling et~al.}(2006)\citenamefont{Kling, Siedschlag,
  Verhoef, Khan, Schultze, Uphues, Ni, Uiberacker, Drescher, Krausz
  et~al.}}]{klin06}
\bibinfo{author}{\bibfnamefont{M.~F.} \bibnamefont{Kling}},
  \bibinfo{author}{\bibfnamefont{C.}~\bibnamefont{Siedschlag}},
  \bibinfo{author}{\bibfnamefont{A.~J.} \bibnamefont{Verhoef}},
  \bibinfo{author}{\bibfnamefont{J.~I.} \bibnamefont{Khan}},
  \bibinfo{author}{\bibfnamefont{M.}~\bibnamefont{Schultze}},
  \bibinfo{author}{\bibfnamefont{T.}~\bibnamefont{Uphues}},
  \bibinfo{author}{\bibfnamefont{Y.}~\bibnamefont{Ni}},
  \bibinfo{author}{\bibfnamefont{M.}~\bibnamefont{Uiberacker}},
  \bibinfo{author}{\bibfnamefont{M.}~\bibnamefont{Drescher}},
  \bibinfo{author}{\bibfnamefont{F.}~\bibnamefont{Krausz}},
  \bibnamefont{et~al.}, \bibinfo{journal}{Science}
  \textbf{\bibinfo{volume}{312}}, \bibinfo{pages}{246} (\bibinfo{year}{2006}).

\bibitem[{\citenamefont{Tong et~al.}(2003)\citenamefont{Tong, Zhao, and
  Lin}}]{tong03}
\bibinfo{author}{\bibfnamefont{X.~M.} \bibnamefont{Tong}},
  \bibinfo{author}{\bibfnamefont{Z.~X.} \bibnamefont{Zhao}}, \bibnamefont{and}
  \bibinfo{author}{\bibfnamefont{C.~D.} \bibnamefont{Lin}},
  \bibinfo{journal}{Phys. Rev. Lett.} \textbf{\bibinfo{volume}{91}},
  \bibinfo{pages}{233203} (\bibinfo{year}{2003}).

\bibitem[{\citenamefont{Fittinghoff et~al.}(1992)\citenamefont{Fittinghoff,
  Bolton, Chang, and Kulander}}]{fitti92}
\bibinfo{author}{\bibfnamefont{D.~N.} \bibnamefont{Fittinghoff}},
  \bibinfo{author}{\bibfnamefont{P.~R.} \bibnamefont{Bolton}},
  \bibinfo{author}{\bibfnamefont{B.}~\bibnamefont{Chang}}, \bibnamefont{and}
  \bibinfo{author}{\bibfnamefont{K.~C.} \bibnamefont{Kulander}},
  \bibinfo{journal}{Phys.~Rev.~Lett.} \textbf{\bibinfo{volume}{69}},
  \bibinfo{pages}{2642} (\bibinfo{year}{1992}).

\bibitem[{\citenamefont{Kondo et~al.}(1993)\citenamefont{Kondo, Sagisaka,
  Tamida, Nabekawa, and Watanabe}}]{kodo93}
\bibinfo{author}{\bibfnamefont{K.}~\bibnamefont{Kondo}},
  \bibinfo{author}{\bibfnamefont{A.}~\bibnamefont{Sagisaka}},
  \bibinfo{author}{\bibfnamefont{T.}~\bibnamefont{Tamida}},
  \bibinfo{author}{\bibfnamefont{Y.}~\bibnamefont{Nabekawa}}, \bibnamefont{and}
  \bibinfo{author}{\bibfnamefont{S.}~\bibnamefont{Watanabe}},
  \bibinfo{journal}{Phys.~Rev.~A} \textbf{\bibinfo{volume}{48}},
  \bibinfo{pages}{R2531} (\bibinfo{year}{1993}).

\bibitem[{\citenamefont{Walker et~al.}(1994)\citenamefont{Walker, Sheehy,
  DiMauro, Agostini, Schafer, and Kulander}}]{walk94}
\bibinfo{author}{\bibfnamefont{B.}~\bibnamefont{Walker}},
  \bibinfo{author}{\bibfnamefont{B.}~\bibnamefont{Sheehy}},
  \bibinfo{author}{\bibfnamefont{L.~F.} \bibnamefont{DiMauro}},
  \bibinfo{author}{\bibfnamefont{P.}~\bibnamefont{Agostini}},
  \bibinfo{author}{\bibfnamefont{K.~J.} \bibnamefont{Schafer}},
  \bibnamefont{and} \bibinfo{author}{\bibfnamefont{K.~C.}
  \bibnamefont{Kulander}}, \bibinfo{journal}{Phys.~Rev.~Lett.}
  \textbf{\bibinfo{volume}{73}}, \bibinfo{pages}{1227} (\bibinfo{year}{1994}).

\bibitem[{\citenamefont{Larochelle et~al.}(1998)\citenamefont{Larochelle,
  Talebpour, and Chin}}]{laro98}
\bibinfo{author}{\bibfnamefont{S.}~\bibnamefont{Larochelle}},
  \bibinfo{author}{\bibfnamefont{A.}~\bibnamefont{Talebpour}},
  \bibnamefont{and} \bibinfo{author}{\bibfnamefont{S.~L.} \bibnamefont{Chin}},
  \bibinfo{journal}{J.~Phys.~B.} \textbf{\bibinfo{volume}{31}},
  \bibinfo{pages}{1201} (\bibinfo{year}{1998}).

\bibitem[{\citenamefont{Weber et~al.}(2000)\citenamefont{Weber, Giessen,
  Weckenbrock, Urbasch, Staudte, Spielberger, Jagutzki, Mergel, Vollmer, and
  D\"orner}}]{webe00_2}
\bibinfo{author}{\bibfnamefont{T.}~\bibnamefont{Weber}},
  \bibinfo{author}{\bibfnamefont{H.}~\bibnamefont{Giessen}},
  \bibinfo{author}{\bibfnamefont{M.}~\bibnamefont{Weckenbrock}},
  \bibinfo{author}{\bibfnamefont{G.}~\bibnamefont{Urbasch}},
  \bibinfo{author}{\bibfnamefont{A.}~\bibnamefont{Staudte}},
  \bibinfo{author}{\bibfnamefont{L.}~\bibnamefont{Spielberger}},
  \bibinfo{author}{\bibfnamefont{O.}~\bibnamefont{Jagutzki}},
  \bibinfo{author}{\bibfnamefont{V.}~\bibnamefont{Mergel}},
  \bibinfo{author}{\bibfnamefont{M.}~\bibnamefont{Vollmer}}, \bibnamefont{and}
  \bibinfo{author}{\bibfnamefont{R.}~\bibnamefont{D\"orner}},
  \bibinfo{journal}{Nature} \textbf{\bibinfo{volume}{405}},
  \bibinfo{pages}{658} (\bibinfo{year}{2000}).

\bibitem[{\citenamefont{Cornaggia and Hering}(2000)}]{corn00}
\bibinfo{author}{\bibfnamefont{C.}~\bibnamefont{Cornaggia}} \bibnamefont{and}
  \bibinfo{author}{\bibfnamefont{P.}~\bibnamefont{Hering}},
  \bibinfo{journal}{Phys.~Rev.~A} \textbf{\bibinfo{volume}{62}},
  \bibinfo{pages}{023403} (\bibinfo{year}{2000}).

\bibitem[{\citenamefont{Guo and Gibson}(2001)}]{guo01}
\bibinfo{author}{\bibfnamefont{C.}~\bibnamefont{Guo}} \bibnamefont{and}
  \bibinfo{author}{\bibfnamefont{G.~N.} \bibnamefont{Gibson}},
  \bibinfo{journal}{Phys.~Rev.~A} \textbf{\bibinfo{volume}{63}},
  \bibinfo{pages}{040701} (\bibinfo{year}{2001}).

\bibitem[{\citenamefont{DeWitt et~al.}(2001)\citenamefont{DeWitt, Wells, and
  Jones}}]{dewi01}
\bibinfo{author}{\bibfnamefont{M.~J.} \bibnamefont{DeWitt}},
  \bibinfo{author}{\bibfnamefont{E.}~\bibnamefont{Wells}}, \bibnamefont{and}
  \bibinfo{author}{\bibfnamefont{R.~R.} \bibnamefont{Jones}},
  \bibinfo{journal}{Phys.~Rev.~Lett.} \textbf{\bibinfo{volume}{87}},
  \bibinfo{pages}{153001} (\bibinfo{year}{2001}).

\bibitem[{\citenamefont{Rudati et~al.}(2004)\citenamefont{Rudati, Chaloupka,
  Agostini, Kulander, and DiMauro}}]{ruda04}
\bibinfo{author}{\bibfnamefont{J.}~\bibnamefont{Rudati}},
  \bibinfo{author}{\bibfnamefont{J.~L.} \bibnamefont{Chaloupka}},
  \bibinfo{author}{\bibfnamefont{P.}~\bibnamefont{Agostini}},
  \bibinfo{author}{\bibfnamefont{K.~C.} \bibnamefont{Kulander}},
  \bibnamefont{and} \bibinfo{author}{\bibfnamefont{L.~F.}
  \bibnamefont{DiMauro}}, \bibinfo{journal}{Phys.~Rev.~Lett.}
  \textbf{\bibinfo{volume}{92}}, \bibinfo{pages}{203001}
  (\bibinfo{year}{2004}).

\bibitem[{\citenamefont{Becker and Faisal}(1996)}]{beck96}
\bibinfo{author}{\bibfnamefont{A.}~\bibnamefont{Becker}} \bibnamefont{and}
  \bibinfo{author}{\bibfnamefont{F.~H.~M.} \bibnamefont{Faisal}},
  \bibinfo{journal}{J.~Phys.~B.} \textbf{\bibinfo{volume}{29}},
  \bibinfo{pages}{L197} (\bibinfo{year}{1996}).

\bibitem[{\citenamefont{Watson et~al.}(1997)\citenamefont{Watson, Sanpera,
  Lappas, Knight, and Burnett}}]{wats97}
\bibinfo{author}{\bibfnamefont{J.~B.} \bibnamefont{Watson}},
  \bibinfo{author}{\bibfnamefont{A.}~\bibnamefont{Sanpera}},
  \bibinfo{author}{\bibfnamefont{D.~G.} \bibnamefont{Lappas}},
  \bibinfo{author}{\bibfnamefont{P.~L.} \bibnamefont{Knight}},
  \bibnamefont{and} \bibinfo{author}{\bibfnamefont{K.}~\bibnamefont{Burnett}},
  \bibinfo{journal}{Phys.~Rev.~Lett.} \textbf{\bibinfo{volume}{78}},
  \bibinfo{pages}{1884} (\bibinfo{year}{1997}).

\bibitem[{\citenamefont{Lappas and van Leeuwen}(1998)}]{lapp98}
\bibinfo{author}{\bibfnamefont{D.~G.} \bibnamefont{Lappas}} \bibnamefont{and}
  \bibinfo{author}{\bibfnamefont{R.}~\bibnamefont{van Leeuwen}},
  \bibinfo{journal}{J.~Phys.~B.} \textbf{\bibinfo{volume}{31}},
  \bibinfo{pages}{L249} (\bibinfo{year}{1998}).

\bibitem[{\citenamefont{Panfili and Liu}(2003)}]{panf03}
\bibinfo{author}{\bibfnamefont{R.}~\bibnamefont{Panfili}} \bibnamefont{and}
  \bibinfo{author}{\bibfnamefont{W.-C.} \bibnamefont{Liu}},
  \bibinfo{journal}{Phys.~Rev.~A} \textbf{\bibinfo{volume}{67}},
  \bibinfo{pages}{043402} (\bibinfo{year}{2003}).

\bibitem[{\citenamefont{Baier et~al.}(2006)\citenamefont{Baier, Ruiz, Plaja,
  and Becker}}]{Baie06}
\bibinfo{author}{\bibfnamefont{S.}~\bibnamefont{Baier}},
  \bibinfo{author}{\bibfnamefont{C.}~\bibnamefont{Ruiz}},
  \bibinfo{author}{\bibfnamefont{L.}~\bibnamefont{Plaja}}, \bibnamefont{and}
  \bibinfo{author}{\bibfnamefont{A.}~\bibnamefont{Becker}},
  \bibinfo{journal}{Phys.~Rev.~A} \textbf{\bibinfo{volume}{74}},
  \bibinfo{pages}{033405} (\bibinfo{year}{2006}).

\bibitem[{\citenamefont{Corkum}(1994)}]{Cork93}
\bibinfo{author}{\bibfnamefont{P.~B.} \bibnamefont{Corkum}},
  \bibinfo{journal}{Phys.~Rev.~Lett.} \textbf{\bibinfo{volume}{71}},
  \bibinfo{pages}{1994} (\bibinfo{year}{1994}).

\bibitem[{\citenamefont{Schafer et~al.}(1993)\citenamefont{Schafer, Yang,
  DiMauro, and Kulander}}]{scha93}
\bibinfo{author}{\bibfnamefont{K.~J.} \bibnamefont{Schafer}},
  \bibinfo{author}{\bibfnamefont{B.}~\bibnamefont{Yang}},
  \bibinfo{author}{\bibfnamefont{L.~F.} \bibnamefont{DiMauro}},
  \bibnamefont{and} \bibinfo{author}{\bibfnamefont{K.~C.}
  \bibnamefont{Kulander}}, \bibinfo{journal}{Phys. Rev. Lett.}
  \textbf{\bibinfo{volume}{70}}, \bibinfo{pages}{1599} (\bibinfo{year}{1993}).

\bibitem[{\citenamefont{Kopold et~al.}(2000)\citenamefont{Kopold, Becker,
  Rottke, and Sandner}}]{kopo00}
\bibinfo{author}{\bibfnamefont{R.}~\bibnamefont{Kopold}},
  \bibinfo{author}{\bibfnamefont{W.}~\bibnamefont{Becker}},
  \bibinfo{author}{\bibfnamefont{H.}~\bibnamefont{Rottke}}, \bibnamefont{and}
  \bibinfo{author}{\bibfnamefont{W.}~\bibnamefont{Sandner}},
  \bibinfo{journal}{Phys.~Rev.~Lett.} \textbf{\bibinfo{volume}{85}},
  \bibinfo{pages}{3781} (\bibinfo{year}{2000}).

\bibitem[{\citenamefont{Lein et~al.}(2000)\citenamefont{Lein, Gross, and
  Engel}}]{lein00}
\bibinfo{author}{\bibfnamefont{M.}~\bibnamefont{Lein}},
  \bibinfo{author}{\bibfnamefont{E.~K.~U.} \bibnamefont{Gross}},
  \bibnamefont{and} \bibinfo{author}{\bibfnamefont{V.}~\bibnamefont{Engel}},
  \bibinfo{journal}{Phys.~Rev.~Lett.} \textbf{\bibinfo{volume}{85}},
  \bibinfo{pages}{4707} (\bibinfo{year}{2000}).

\bibitem[{\citenamefont{Sacha and Eckhardt}(2001)}]{sach01}
\bibinfo{author}{\bibfnamefont{K.}~\bibnamefont{Sacha}} \bibnamefont{and}
  \bibinfo{author}{\bibfnamefont{B.}~\bibnamefont{Eckhardt}},
  \bibinfo{journal}{Phys.~Rev.~A} \textbf{\bibinfo{volume}{63}},
  \bibinfo{pages}{043414} (\bibinfo{year}{2001}).

\bibitem[{\citenamefont{Fu et~al.}(2001)\citenamefont{Fu, Liu, Chen, and
  Chen}}]{fu01}
\bibinfo{author}{\bibfnamefont{L.-B.} \bibnamefont{Fu}},
  \bibinfo{author}{\bibfnamefont{J.}~\bibnamefont{Liu}},
  \bibinfo{author}{\bibfnamefont{J.}~\bibnamefont{Chen}}, \bibnamefont{and}
  \bibinfo{author}{\bibfnamefont{S.-G.} \bibnamefont{Chen}},
  \bibinfo{journal}{Phys.~Rev.~A} \textbf{\bibinfo{volume}{63}},
  \bibinfo{pages}{043416} (\bibinfo{year}{2001}).

\bibitem[{\citenamefont{Panfili et~al.}(2001)\citenamefont{Panfili, Eberly, and
  Haan}}]{panf01}
\bibinfo{author}{\bibfnamefont{R.}~\bibnamefont{Panfili}},
  \bibinfo{author}{\bibfnamefont{J.~H.} \bibnamefont{Eberly}},
  \bibnamefont{and} \bibinfo{author}{\bibfnamefont{S.~L.} \bibnamefont{Haan}},
  \bibinfo{journal}{Optics Express} \textbf{\bibinfo{volume}{8}},
  \bibinfo{pages}{431} (\bibinfo{year}{2001}).

\bibitem[{\citenamefont{Barna and Rost}(2003)}]{barn03}
\bibinfo{author}{\bibfnamefont{I.~F.} \bibnamefont{Barna}} \bibnamefont{and}
  \bibinfo{author}{\bibfnamefont{J.~M.} \bibnamefont{Rost}},
  \bibinfo{journal}{Eur. Phys. J. D} \textbf{\bibinfo{volume}{27}},
  \bibinfo{pages}{287} (\bibinfo{year}{2003}).

\bibitem[{\citenamefont{Colgan et~al.}(2004)\citenamefont{Colgan, Pindzola, and
  Robicheaux}}]{colg04}
\bibinfo{author}{\bibfnamefont{J.}~\bibnamefont{Colgan}},
  \bibinfo{author}{\bibfnamefont{M.~S.} \bibnamefont{Pindzola}},
  \bibnamefont{and}
  \bibinfo{author}{\bibfnamefont{F.}~\bibnamefont{Robicheaux}},
  \bibinfo{journal}{Phys. Rev. Lett.} \textbf{\bibinfo{volume}{93}},
  \bibinfo{pages}{053201} (\bibinfo{year}{2004}).

\bibitem[{\citenamefont{Ho et~al.}(2005)\citenamefont{Ho, Panfili, Haan, and
  Eberly}}]{ho05_1}
\bibinfo{author}{\bibfnamefont{P.~J.} \bibnamefont{Ho}},
  \bibinfo{author}{\bibfnamefont{R.}~\bibnamefont{Panfili}},
  \bibinfo{author}{\bibfnamefont{S.~L.} \bibnamefont{Haan}}, \bibnamefont{and}
  \bibinfo{author}{\bibfnamefont{J.~H.} \bibnamefont{Eberly}},
  \bibinfo{journal}{Phys.~Rev.~Lett.} \textbf{\bibinfo{volume}{94}},
  \bibinfo{pages}{093002} (\bibinfo{year}{2005}).

\bibitem[{\citenamefont{Ho and Eberly}(2005)}]{ho05_2}
\bibinfo{author}{\bibfnamefont{P.~J.} \bibnamefont{Ho}} \bibnamefont{and}
  \bibinfo{author}{\bibfnamefont{J.~H.} \bibnamefont{Eberly}},
  \bibinfo{journal}{Phys.~Rev.~Lett.} \textbf{\bibinfo{volume}{95}},
  \bibinfo{pages}{193002} (\bibinfo{year}{2005}).

\bibitem[{\citenamefont{Ruiz et~al.}(2005)\citenamefont{Ruiz, Plaja, and
  Roso}}]{ruiz05}
\bibinfo{author}{\bibfnamefont{C.}~\bibnamefont{Ruiz}},
  \bibinfo{author}{\bibfnamefont{L.}~\bibnamefont{Plaja}}, \bibnamefont{and}
  \bibinfo{author}{\bibfnamefont{L.}~\bibnamefont{Roso}},
  \bibinfo{journal}{Phys.~Rev.~Lett.} \textbf{\bibinfo{volume}{94}},
  \bibinfo{pages}{063002} (\bibinfo{year}{2005}).

\bibitem[{\citenamefont{Horner et~al.}(2007)\citenamefont{Horner, Morales,
  Rescigno, Mart\'{\i}n, and McCurdy}}]{horn07}
\bibinfo{author}{\bibfnamefont{D.~A.} \bibnamefont{Horner}},
  \bibinfo{author}{\bibfnamefont{F.}~\bibnamefont{Morales}},
  \bibinfo{author}{\bibfnamefont{T.~N.} \bibnamefont{Rescigno}},
  \bibinfo{author}{\bibfnamefont{F.}~\bibnamefont{Mart\'{\i}n}},
  \bibnamefont{and} \bibinfo{author}{\bibfnamefont{C.~W.}
  \bibnamefont{McCurdy}}, \bibinfo{journal}{Phys.~Rev.~A}
  \textbf{\bibinfo{volume}{76}}, \bibinfo{pages}{030701(R)}
  (\bibinfo{year}{2007}).

\bibitem[{\citenamefont{Prauzner-Bechcicki
  et~al.}(2007)\citenamefont{Prauzner-Bechcicki, Sacha, Eckhardt, and
  Zakrzewski}}]{prau07}
\bibinfo{author}{\bibfnamefont{J.~S.} \bibnamefont{Prauzner-Bechcicki}},
  \bibinfo{author}{\bibfnamefont{K.}~\bibnamefont{Sacha}},
  \bibinfo{author}{\bibfnamefont{B.}~\bibnamefont{Eckhardt}}, \bibnamefont{and}
  \bibinfo{author}{\bibfnamefont{J.}~\bibnamefont{Zakrzewski}},
  \bibinfo{journal}{Phys.~Rev.~Lett.} \textbf{\bibinfo{volume}{98}},
  \bibinfo{pages}{203002} (\bibinfo{year}{2007}).

\bibitem[{\citenamefont{Feist et~al.}(2008)\citenamefont{Feist, Nagele,
  Pazourek, Persson, Schneider, Collins, and Burgd\"orfer}}]{feis08}
\bibinfo{author}{\bibfnamefont{J.}~\bibnamefont{Feist}},
  \bibinfo{author}{\bibfnamefont{S.}~\bibnamefont{Nagele}},
  \bibinfo{author}{\bibfnamefont{R.}~\bibnamefont{Pazourek}},
  \bibinfo{author}{\bibfnamefont{E.}~\bibnamefont{Persson}},
  \bibinfo{author}{\bibfnamefont{B.~I.} \bibnamefont{Schneider}},
  \bibinfo{author}{\bibfnamefont{L.~A.} \bibnamefont{Collins}},
  \bibnamefont{and}
  \bibinfo{author}{\bibfnamefont{J.}~\bibnamefont{Burgd\"orfer}},
  \bibinfo{journal}{Phys.~Rev.~A} \textbf{\bibinfo{volume}{77}},
  \bibinfo{pages}{043420} (\bibinfo{year}{2008}).

\bibitem[{\citenamefont{Bryan et~al.}(2006)\citenamefont{Bryan, Stebbings,
  McKenna, English, Suresh, Wood, Srigengan, Turcu, Smith, Divall
  et~al.}}]{brya06}
\bibinfo{author}{\bibfnamefont{W.~A.} \bibnamefont{Bryan}},
  \bibinfo{author}{\bibfnamefont{S.~L.} \bibnamefont{Stebbings}},
  \bibinfo{author}{\bibfnamefont{J.}~\bibnamefont{McKenna}},
  \bibinfo{author}{\bibfnamefont{E.~M.~L.} \bibnamefont{English}},
  \bibinfo{author}{\bibfnamefont{M.}~\bibnamefont{Suresh}},
  \bibinfo{author}{\bibfnamefont{J.}~\bibnamefont{Wood}},
  \bibinfo{author}{\bibfnamefont{B.}~\bibnamefont{Srigengan}},
  \bibinfo{author}{\bibfnamefont{I.~C.~E.} \bibnamefont{Turcu}},
  \bibinfo{author}{\bibfnamefont{J.~M.} \bibnamefont{Smith}},
  \bibinfo{author}{\bibfnamefont{E.~J.} \bibnamefont{Divall}},
  \bibnamefont{et~al.}, \bibinfo{journal}{Nature Physics}
  \textbf{\bibinfo{volume}{2}}, \bibinfo{pages}{379} (\bibinfo{year}{2006}).

\bibitem[{\citenamefont{Panfili et~al.}(2002)\citenamefont{Panfili, Haan, and
  Eberly}}]{panf02}
\bibinfo{author}{\bibfnamefont{R.}~\bibnamefont{Panfili}},
  \bibinfo{author}{\bibfnamefont{S.~L.} \bibnamefont{Haan}}, \bibnamefont{and}
  \bibinfo{author}{\bibfnamefont{J.~H.} \bibnamefont{Eberly}},
  \bibinfo{journal}{Phys.~Rev.~Lett.} \textbf{\bibinfo{volume}{89}},
  \bibinfo{pages}{113001} (\bibinfo{year}{2002}).

\bibitem[{\citenamefont{Liu et~al.}(2007)\citenamefont{Liu, Ye, Chen, and
  Liu}}]{liu07}
\bibinfo{author}{\bibfnamefont{J.}~\bibnamefont{Liu}},
  \bibinfo{author}{\bibfnamefont{D.~F.} \bibnamefont{Ye}},
  \bibinfo{author}{\bibfnamefont{J.}~\bibnamefont{Chen}}, \bibnamefont{and}
  \bibinfo{author}{\bibfnamefont{X.}~\bibnamefont{Liu}},
  \bibinfo{journal}{Phys.~Rev.~Lett.} \textbf{\bibinfo{volume}{99}},
  \bibinfo{pages}{013003} (\bibinfo{year}{2007}).

\bibitem[{\citenamefont{Mauger et~al.}(2009)\citenamefont{Mauger, Chandre, and
  Uzer}}]{maug09}
\bibinfo{author}{\bibfnamefont{F.}~\bibnamefont{Mauger}},
  \bibinfo{author}{\bibfnamefont{C.}~\bibnamefont{Chandre}}, \bibnamefont{and}
  \bibinfo{author}{\bibfnamefont{T.}~\bibnamefont{Uzer}},
  \bibinfo{journal}{Phys. Rev. Lett.} \textbf{\bibinfo{volume}{102}},
  \bibinfo{pages}{173002} (\bibinfo{year}{2009}).

\bibitem[{\citenamefont{Prauzner-Bechcicki
  et~al.}(2005)\citenamefont{Prauzner-Bechcicki, Sacha, Eckhardt, and
  Zakrzewski}}]{Prau05}
\bibinfo{author}{\bibfnamefont{J.~S.} \bibnamefont{Prauzner-Bechcicki}},
  \bibinfo{author}{\bibfnamefont{K.}~\bibnamefont{Sacha}},
  \bibinfo{author}{\bibfnamefont{B.}~\bibnamefont{Eckhardt}}, \bibnamefont{and}
  \bibinfo{author}{\bibfnamefont{J.}~\bibnamefont{Zakrzewski}},
  \bibinfo{journal}{Phys.~Rev.~A} \textbf{\bibinfo{volume}{71}},
  \bibinfo{pages}{033407} (\bibinfo{year}{2005}).

\bibitem[{\citenamefont{Lein et~al.}(2002)\citenamefont{Lein, Kreibich, Gross,
  and Engel}}]{Lein02}
\bibinfo{author}{\bibfnamefont{M.}~\bibnamefont{Lein}},
  \bibinfo{author}{\bibfnamefont{T.}~\bibnamefont{Kreibich}},
  \bibinfo{author}{\bibfnamefont{E.~K.~U.} \bibnamefont{Gross}},
  \bibnamefont{and} \bibinfo{author}{\bibfnamefont{V.}~\bibnamefont{Engel}},
  \bibinfo{journal}{Phys.~Rev.~A} \textbf{\bibinfo{volume}{65}},
  \bibinfo{pages}{033403} (\bibinfo{year}{2002}).

\bibitem[{\citenamefont{Saugout et~al.}(2008)\citenamefont{Saugout, Charron,
  and Cornaggia}}]{Saug08}
\bibinfo{author}{\bibfnamefont{S.}~\bibnamefont{Saugout}},
  \bibinfo{author}{\bibfnamefont{E.}~\bibnamefont{Charron}}, \bibnamefont{and}
  \bibinfo{author}{\bibfnamefont{C.}~\bibnamefont{Cornaggia}},
  \bibinfo{journal}{Phys.~Rev.~A} \textbf{\bibinfo{volume}{77}},
  \bibinfo{pages}{023404} (\bibinfo{year}{2008}).

\bibitem[{\citenamefont{Cvitanovi\'c et~al.}(2008)\citenamefont{Cvitanovi\'c,
  Artuso, Mainieri, Tanner, and Vattay}}]{chaosbook}
\bibinfo{author}{\bibfnamefont{P.}~\bibnamefont{Cvitanovi\'c}},
  \bibinfo{author}{\bibfnamefont{R.}~\bibnamefont{Artuso}},
  \bibinfo{author}{\bibfnamefont{R.}~\bibnamefont{Mainieri}},
  \bibinfo{author}{\bibfnamefont{G.}~\bibnamefont{Tanner}}, \bibnamefont{and}
  \bibinfo{author}{\bibfnamefont{G.}~\bibnamefont{Vattay}},
  \emph{\bibinfo{title}{Chaos: Classical and Quantum}}
  (\bibinfo{publisher}{Niels Bohr Institute}, \bibinfo{address}{Copenhagen},
  \bibinfo{year}{2008}), \bibinfo{note}{{\tt
  {http://ChaosBook.org}{ChaosBook.org}}}.

\bibitem[{\citenamefont{Shchekinova et~al.}(2006)\citenamefont{Shchekinova,
  Chandre, and Uzer}}]{Shch06}
\bibinfo{author}{\bibfnamefont{E.}~\bibnamefont{Shchekinova}},
  \bibinfo{author}{\bibfnamefont{C.}~\bibnamefont{Chandre}}, \bibnamefont{and}
  \bibinfo{author}{\bibfnamefont{T.}~\bibnamefont{Uzer}},
  \bibinfo{journal}{Phys.~Rev.~A} \textbf{\bibinfo{volume}{74}},
  \bibinfo{pages}{043417} (\bibinfo{year}{2006}).

\bibitem[{\citenamefont{Greene}(1979)}]{gree79}
\bibinfo{author}{\bibfnamefont{J.~M.} \bibnamefont{Greene}},
  \bibinfo{journal}{J. Math. Phys.} \textbf{\bibinfo{volume}{20}},
  \bibinfo{pages}{1183} (\bibinfo{year}{1979}).

\bibitem[{\citenamefont{Bandrauk et~al.}(2005)\citenamefont{Bandrauk,
  Chelkowski, and Goudreau}}]{Band05}
\bibinfo{author}{\bibfnamefont{A.~D.} \bibnamefont{Bandrauk}},
  \bibinfo{author}{\bibfnamefont{S.}~\bibnamefont{Chelkowski}},
  \bibnamefont{and} \bibinfo{author}{\bibfnamefont{S.}~\bibnamefont{Goudreau}},
  \bibinfo{journal}{J. Mod. Opt.} \textbf{\bibinfo{volume}{52}},
  \bibinfo{pages}{411} (\bibinfo{year}{2005}).

\end{thebibliography}

\end{document}